\documentclass[showpacs,amssymb,10pt,reprint,aps,prd,nofootinbib,floatfix,superscriptaddress]{revtex4-2}
\usepackage{graphicx,epsfig,amssymb} 
\usepackage{appendix}
\usepackage{relsize}
\usepackage{amsmath,amsfonts, times}
\usepackage{bm} 
\usepackage[normalem]{ulem}
\usepackage{epstopdf}
\usepackage[caption=false]{subfig}
\usepackage[usenames]{color} 
\usepackage{mathrsfs}
\usepackage{tensor}
\usepackage{natbib}
\usepackage{soul}
\usepackage{subfig}
\usepackage[utf8x]{inputenc}
\usepackage{tikz}
\usetikzlibrary{decorations.pathmorphing}

\definecolor{coolblack}{rgb}{0.0, 0.18, 0.39}
\definecolor{darkred}{rgb}{0.5,0,0}
\definecolor{darkgreen}{rgb}{0,0.5,0}
\definecolor{darkblue}{rgb}{0,0,0.5}
\definecolor{lapislazuli}{rgb}{0.15, 0.38, 0.61}
\definecolor{venetianred}{rgb}{0.78, 0.03, 0.08}
\definecolor{bleudefrance}{rgb}{0.19, 0.55, 0.91}
\definecolor{dogwoodrose}{rgb}{0.84, 0.09, 0.41}

\begin{document}
 \title{Echoes from bounded universes}
	\author{Renan B. Magalh\~aes}
	\email{renan.magalhaes@icen.ufpa.br}
	\affiliation{Programa de P\'os-Gradua\c{c}\~{a}o em F\'{\i}sica, Universidade 
		Federal do Par\'a, 66075-110, Bel\'em, Par\'a, Brazil.}
	\affiliation{Departamento de F{\'i}sica Te{\'o}rica and \textit{IFIC}, Centro Mixto Universidad de Valencia - \textit{CSIC}. Universidad de Valencia, Burjassot-46100, Valencia, Spain.}	
	\author{Andreu Mas{\'o}-Ferrando}
	\email{andreu.maso@uv.es}
	\affiliation{Departamento de F{\'i}sica Te{\'o}rica and \textit{IFIC}, Centro Mixto Universidad de Valencia - \textit{CSIC}. Universidad de Valencia, Burjassot-46100, Valencia, Spain.}
		\author{Flavio Bombacigno}
	\email{flavio2.bombacigno@uv.es}
	\affiliation{Departamento de F{\'i}sica Te{\'o}rica and \textit{IFIC}, Centro Mixto Universidad de Valencia - \textit{CSIC}. Universidad de Valencia, Burjassot-46100, Valencia, Spain.}	
	\author{Gonzalo J. Olmo}
	\email{gonzalo.olmo@uv.es}
	\affiliation{Departamento de F{\'i}sica Te{\'o}rica and \textit{IFIC}, Centro Mixto Universidad de Valencia - \textit{CSIC}. Universidad de Valencia, Burjassot-46100, Valencia, Spain.}
 \affiliation{Departamento de F\'isica, Universidade Federal do Cear\'a, Caixa Postal 6030, Campus do Pici, 60455-760 Fortaleza, Cear\'a, Brazil}
	\author{Lu\'is C. B. Crispino}
	\email{crispino@ufpa.br}
\affiliation{Programa de P\'os-Gradua\c{c}\~{a}o em F\'{\i}sica, Universidade Federal do Par\'a, 66075-110, Bel\'em, Par\'a, Brazil.}%
\affiliation{Departamento de Matem\'atica da Universidade de Aveiro and Centre for Research and Development  in Mathematics and Applications (CIDMA), Campus de Santiago, 3810-183 Aveiro, Portugal.}
	
	\begin{abstract}
We construct a general class of modified Ellis wormholes, where one asymptotic Minkowski region is replaced by a bounded 2-sphere core, characterized by asymptotic finite areal radius. We pursue an in-depth analysis of the resulting geometry, outlining that geodesic completeness is guaranteed also when the radial function asymptotically shrinks to zero. Then, we study the evolution of scalar perturbations, bringing out how these geometric configurations can in principle affect the time-domain profiles of quasinormal modes, pointing out the distinctive features with respect to other black holes or wormholes geometries.
\end{abstract}
	\date{\today}
	\maketitle
\section{Introduction}\label{sec:int}
The advent of gravitational waves astronomy \cite{LIGOScientific:2016aoc} and the technical progress made in very large base-line interferometry \cite{EventHorizonTelescope:2019dse} is enabling us to confront predictions of gravitational theories in the strong field regime with observational data. In this sense, significant theoretical efforts are being devoted to the search and study of exotic compact objects that may offer a reasonable alternative to the black hole paradigm of Einstein’s theory of General Relativity (GR). Such objects could lead to effects not expected in GR and, for this reason, their analysis could help identify observational signatures of new physics. Among the most popular alternatives, one finds hairy or regularized black holes \cite{Herdeiro:2014goa,Herdeiro:2016tmi,Dymnikova:1992ux,Ansoldi:2008jw}, boson stars \cite{Liebling:2012fv,Visinelli:2021uve}, gravastars \cite{mazur2023gravitational}, wormholes \cite{Visser:1995cc}, and black bounces \cite{Simpson:2018tsi}. In this work, we put the focus on a new kind of exotic object closely related with wormholes but with proper observational features that could allow us to tell them apart from the usual wormholes found in the literature. 

In the conventional approach to the resolution of the  field equations of a given gravitational system, one typically begins with a reasonable description of the matter sources and then employs the Einstein field equations to derive the corresponding spacetime geometry. When dealing with wormhole configurations, however, the process is usually inverted.  In fact, once the desired spacetime geometry is selected, it is possible to use Einstein's equations in reverse to determine the matter distribution responsible for it. Despite violating various energy conditions, wormholes still stand out because of their theoretical implications, as they offer valuable insights into the foundational aspects of gravity models and their possible non trivial topological structure. Wormholes are not only relevant for interstellar travel \cite{Morris:1988cz}, which is a well-known and popular application recurrently used in the literature, but could be also crucial to better understand the nature of quantum entanglement \cite{Maldacena:2013xja}. For these reasons, it is important to explicitly address the theoretical implications of their existence and the proper observational signatures that could signal their presence in astrophysical scenarios \cite{Nandi:2006ds,Ohgami:2015nra}.

The emission of gravitational waves by the coalescence of compact objects, for instance, can clearly distinguish between black holes (objects with an event horizon) and wormholes (no event horizon), as the latter are expected to emit a series of echoes~\cite{Cardoso:2016rao} which are not present in the case of objects with a horizon, from which nothing can come out (though these echoes are not unique to wormholes, see for instance \cite{Cardoso:2016oxy}). Also, the distortion of light trajectories by black holes and wormholes might be quite different, not only because wormholes may allow light to come out from their internal region but also because the number and structure of light rings might exhibit different behaviors \cite{Wielgus:2020uqz,guerrero2022light}. These aspects can be used to identify a number of treats that allow to characterize the various specimens present in the zoo of compact objects.

In this sense, it should be noted that among the billions of compact objects present in the universe, the observation of events apparently compatible with our predictions does not rule out the existence of objects of a different nature. Current detectors' sensitivities are still insufficient to break significant degeneracies in the data \cite{CalderonBustillo:2020fyi}, which demands for new generations of observatories able to provide a clearer picture of the nature of such objects \cite{Cardoso:2019rvt}. In the meantime, the existence and consistency of exotic compact objects can be pursued theoretically, with the aim of identifying their proper signatures and potential phenomenological implications.

It is important, therefore, to have a clear understanding of the proper features that define and identify wormholes. From an astrophysical perspective, one would expect that wormholes should be spherical or axisymmetric (rotating). Accordingly, the presence of spherically symmetric wormholes is typically identified in the literature by the existence of a two-sphere of minimal non-zero area, which represents the wormhole throat \cite{Visser:1995cc,Visser:1997yn}. However, the observation of a minimal two-sphere, which is a local characteristic, cannot be used to infer anything about the global properties of the spacetime, such as its topology. Thus, in order to broaden our understanding of their observational features, it is necessary to consider scenarios in which typical local properties of wormholes are present but where their proper global aspects are not. In particular, since in standard wormholes the throat connects two unbounded (typically asymptotically flat) regions, one may consider spacetimes where the throat connects an unbounded region to a bounded one. Spacetimes with this characteristic have been recently found as the end product of the collapse of boson stars in quadratic Palatini gravity \cite{Maso-Ferrando:2023nju,Maso-Ferrando:2023wtz}. In that scenario, the central region of the collapsed object experiences a kind of inflationary stretching that leads to the formation of an expanding bubble (or baby universe) behind the event horizon. Inspired by that phenomenon, here we consider a static setting in which an asymptotically Minkowskian region is attached to a bounded bubble via a minimal two-sphere. The gravitational echoes generated by such bounded universes and their geodesic structure will be scrutinized and compared with those of more standard wormholes. 

One of the simplest wormhole configurations is represented by the well-known Ellis wormhole (EWH) \cite{ellis1973ether}, which is made of two asymptotically flat regions connected by a throat\footnote{In this work we do not focus on wormhole geometries involving Schwarzschild-like deformation (see for instance \cite{Simpson:2018tsi}), which will be the object of a forthcoming analysis \cite{iamthefuture}.}. In this work we introduce the idea that one of the infinite asymptotic region, i.e. the Minkowski spacetime of EWHs, could be possibly replaced by some inner spatial volume endowed with finiteness in one or more directions. The (partial) compactness of this new internal region is thus expected to induce specific observable effects on both the shadow of the object and on its spectroscopic properties, where peculiar absorption effects, resonances, and echoes may arise depending on the typical size of the internal region and the different boundary conditions involved.

In this work we move a first step on in this direction by looking at minimal deformations of the classic EWH \cite{ellis1973ether} considering that one of the asymptotic Minkowski regions is replaced by a bounded 2-sphere with asymptotically constant areal radius. We expect that the finiteness of the radial function in this internal region plays a relevant role in the propagation of perturbations, because the effective potential perceived by scalar modes, hereinafter $V_{\Phi}$, critically depends on the areal radius behavior. In particular, we refer to the possibility that echoes could be generated, bringing the question of how to distinguish, in a phenomenological sense, our geometrical setting from analogous results in different scenarios \cite{cardoso2016gravitational,cardoso2017tests,bueno2018echoes,bronnikov2020echoes,churilova2021wormholes,yang2021echoes}. Moreover, among the possible geometric realizations studied, we will consider the case in which the asymptotic areal radius vanishe, implying that the geodesic completeness of the resulting spacetime is not a priori guaranteed (see \cite{afonso2019new,magalhaes2022compact} for examples of geodesically incomplete wormholes). In order to clarify this delicate issue we analyze to some extent the geodesic motion for radial (analytically) and non-radial (numerically) trajectories, confirming that geodesic completeness is insensitive to the actual value of the asymptotic areal radius, which turns out to affect only the presence or not of inner region bounces for nonradial motion. 

The paper is organized as follows. In sec.~\ref{sec:AWH} we introduce the model and we discuss the general geometric properties, by also presenting a detailed analysis of the energy condition violation; in sec.~\ref{sec:geo} we perform an exhaustive investigation of geodesic motion and photon orbit stability; in sec.~\ref{sec:SFP} scalar perturbations are addressed and we show how the appearance and the properties of echoes are affected by the geometry; in sec.~\ref{sec:conclusion} we report our final comments.

Spacetime signature is chosen mostly plus, i.e. $\eta_{\mu\nu}=\text{diag}(-1,1,1,1)$ and we adopted units in which $8\pi G=1$ and $c=1$.
	
\section{The model}\label{sec:AWH}
 One of the simplest (symmetric) traversable wormholes in the literature is represented by the EWH, whose line element is given by~\cite{ellis1973ether}
\begin{equation}
\label{eq:EWH}
ds^2 = -dt^2+dx^2+r(x)^2(d\theta^2+\sin^2\theta d\varphi^2),
\end{equation}
with radial coordinate $x\in (-\infty,\infty)$ and areal radius $r(x) = \sqrt{x^2+a^2}$. The areal radius has a regular minimum at $x=0$, which corresponds to the wormhole throat with radius $a$. As $|x|$ increases, the areal radius monotonically grows and the line element~\eqref{eq:EWH} approaches Minkowski spacetime as $x\to\pm\infty$.

Here we present a toy model of an asymmetric wormhole-like object, characterized by a bounded 2-sphere radius inner region surrounded by an asymptotically flat exterior, formally described by the line element~\eqref{eq:EWH}, but with the modified radial function
\begin{equation}
 r^2(x)=	\left\{
\begin{array}{ll}
x^2+a^2,&x\geq 0, \\
x^2+a^2-(x^2+a^2-R^2)\tanh^2(c\,x^2),&x<0,
\end{array}\right.\label{eq:areal_radius} 
\end{equation}
where $a$ and $R$ are constants with dimension of length, while $c$ has dimension of $\text{length}^{-2}$. One can check that \eqref{eq:areal_radius} is endowed with a throat-like structure in $x=0$, where $r(x)$ exhibits the regular minimum $r^2(0)=a^2$, with $a$ taken to be the throat radius. In the asymptotic internal region, as $x\to-\infty$, the radial function $r^2(x)\to R^2$, that is, the 2-sphere radius is asymptotically bounded. In particular, when $R^2=0$, the 2-sphere radius shrinks to zero exponentially. The role of the parameter $c$ is to control how much the radial function $r^2(x)$ departs from a parabola close to the throat, so that the asymptotic value $R$ is reached faster as one increases the value of $c$. If $c=0$, the radial function~\eqref{eq:areal_radius} reduces to the one of the EWH, that is, $r(x)^2 = x^2+a^2$. 

\begin{figure}
    \centering
    \includegraphics[width=\columnwidth]{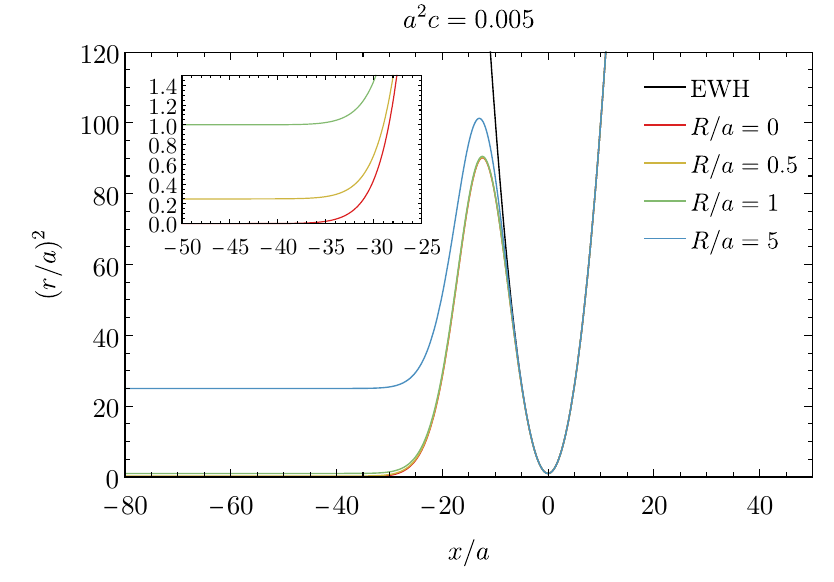}
    \includegraphics[width=\columnwidth]{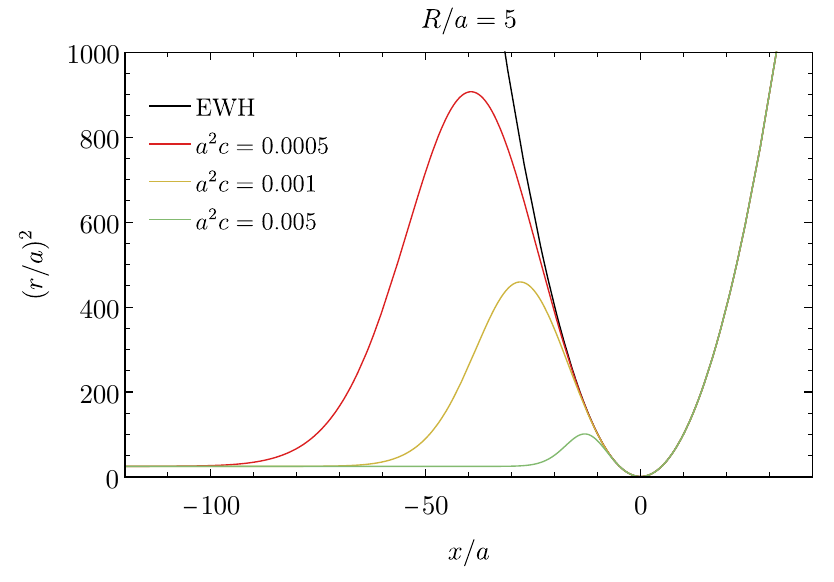}
    \caption{Modified areal radius~\eqref{eq:areal_radius}. Top panel: $r(x)^2$ for fixed value of $c$ and different asymptotic 2-sphere radius $R$. Bottom panel: $r(x)^2$ for fixed $R$ and some choices of $c$. We normalized the plots with the throat radius $a$.}
    \label{fig:areal_radius}
\end{figure}

In this model, the outer universe ($x\geq 0$) is described by the same line element as the EWH spacetime. The modified areal radius~\eqref{eq:areal_radius}, however, sharply modifies the structure of the inner universe ($x<0$). In Fig.~\ref{fig:areal_radius}
we exhibit the behavior of the modified areal radius squared in both sides of the throat and compare it with the standard parabolic behavior of EWH ($r(x)^2=x^2+a^2$). In the top panel we fixed the value of the parameter $c$ and consider some values of the asymptotic 2-sphere radius $R$. In the bottom panel we fixed the asymptotic radius and consider different values of $c$. From Fig.~\ref{fig:areal_radius} one notices that the areal radius may present a local maximum in the inner region. One can check that maximum location, $x_m$, must satisfy $M(x_m)= 0 $ and $r_{xx}(x_m)<0$, where
\begin{equation}
    M(x)\equiv 1 - 2c(x^2+a^2-R^2)\tanh(c\,x^2)
\end{equation}
and $r_{xx}$ stands for the second derivative of $r(x)$ (similarly $r_x$ denotes the first derivative of $r(x)$).
Since $M(x)$ is continuous in the interval $(-\infty,0)$, and $M(x)\to -\infty$ as $x\to-\infty$ and $M(x)\to 1$ as $x\to 0^{-}$, there is at least one point $x_m\in (-\infty,0)$ such that $M(x_m)=0$, and this point is a maximum whether $r_{xx}(x_m)<0$. 

\subsection{Embedding diagrams}
\begin{figure*}
    \centering
    \includegraphics[width=0.9\columnwidth]{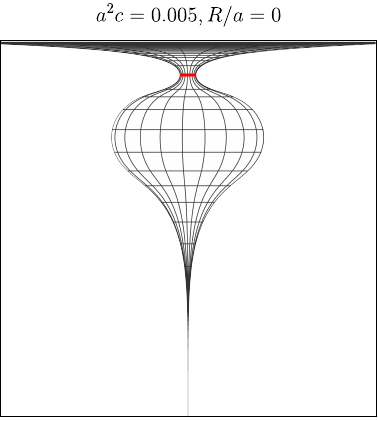}
    \includegraphics[width=0.9\columnwidth]{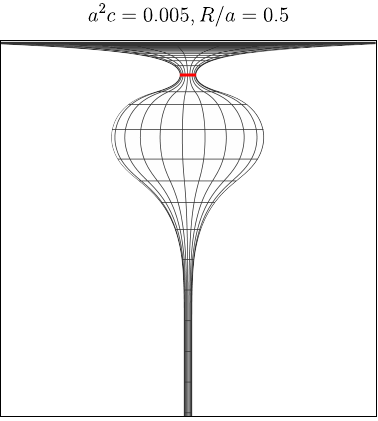}\\\includegraphics[width=0.9\columnwidth]{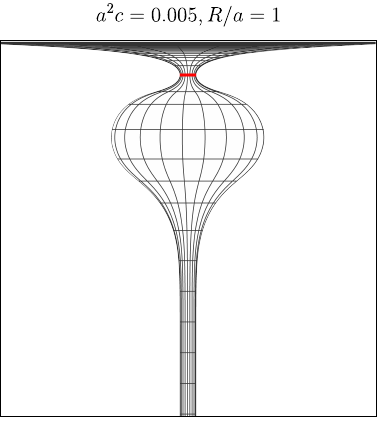}
    \includegraphics[width=0.9\columnwidth]{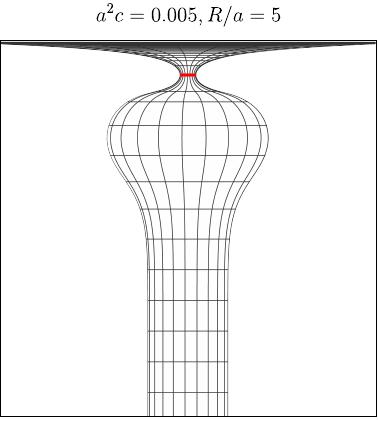}
    \caption{Embedding diagrams of finite 2-sphere radius EWHs with fixed value of $c$. In the top-left panel we exhibit a wormhole-like object in which the 2-sphere radius shrinks to zero, creating a sort of ``bubble'' in the internal region. The other panels correspond to asymptotic 2-sphere radius $R$ smaller (top-right), equal (bottom-left) and bigger (bottom-right) than the wormhole throat $a$ (represented by the red line).}
    \label{fig:embedding_diagrams}
\end{figure*}

The modified areal radius~\eqref{eq:areal_radius}, therefore, significantly changes the structure of the modified EWH internal region. That can be properly visualized by drawing the embedding diagrams of the model for different asymptotic 2-sphere radius. To do so, we consider at a given time coordinate $t$ the equatorial plane $\theta=\pi/2$, where the metric~\eqref{eq:EWH} reduces to $ds^2=dx^2+r(x)^2d\phi^2$. Now, one may embed this hypersurface in the three-dimensional Euclidean space (written in cylindrical coordinates) $ds^2 = dz^2 + d\rho^2 + \rho^2d\phi^2$. By identifying the polar radius with the radial function, i.e. $\rho=r(x)$, one obtains the differential equation
\begin{equation}
    \label{eq:embedding}
    z_x^2 = 1-r_x^2,
\end{equation}
which can be solved for $z(x)$. Together with \eqref{eq:areal_radius}, they can be plotted in the plane $r-z$ to build the embedding diagram as the revolution surface of the curve $\gamma(x)=(r(x),z(x))$ about axis $r=0$. In Fig.~\ref{fig:embedding_diagrams} we show the embedding diagrams of four modified EWHs with bounded 2-sphere radius in the inner region. 
In particular, in the top-left panel, we exhibit the embedding diagram of a wormhole-like object whose 2-sphere radius shrinks to zero in the asymptotic region, creating a sort of ``bubble'' below the throat in the embedding diagram. Additionally, we also show three configurations of finite $R$, with $R<a$, $R=a$ and $R>a$. 

\subsection{Curvature, energy density and pressures}
To have a better physical view of the solutions modeled by the line element~\eqref{eq:EWH} with areal radius~\eqref{eq:areal_radius}, one can compute the curvature invariants related with it, which read

\begin{align}
&g_{\mu\nu}R^{\mu\nu}= \dfrac{2\left(1-r_x^2-2rr_{xx}\right)}{r^2},\\
&R_{\mu\nu}R^{\mu\nu} =\dfrac{2\left(\left(r_x^2-1\right)^2+2r\left(r_x^2-1\right)r_{xx}+3r^2r_{xx}^2\right)}{r^4},\\
&R_{\alpha\beta\gamma\lambda}R^{\alpha\beta\gamma\lambda} =\frac{4 \left(1-2 r_x^2+r_x^4+
   2r^2r_{xx}^2\right)}{r^4}.
\end{align}

Since the spacetime is asymptotically flat in the exterior region all the curvature scalars vanish very far from the throat. In the internal region, instead, curvature invariants may be bounded or unbounded depending on the asymptotic 2-sphere radius $R$. As the radial coordinate approaches $x\to-\infty$, one finds that $g_{\mu\nu}R^{\mu\nu}\to 2/R^2$, $R_{\mu\nu}R^{\mu\nu}\to 2/R^4$ and $R_{\alpha\beta\gamma\lambda}R^{\alpha\beta\gamma\lambda}\to 4/R^4$, corresponding to the curvature invariants of a 2-sphere of radius $R$. If $R=0$ the curvature scalars are unbounded in the internal region, as they diverge as $x\to-\infty$.  However, such a behaviour does not represent actually any pathology in the spacetime, since all the geodesics can be extended forever in this geometry, and no spacetime singularity is present (see discussion in Sec.~\ref{sec:geo}).

It is a well established fact that standard EWHs can be sustained in GR  only in the presence of exotic matter fields, resulting in the explicit violation of the different energy conditions~\cite{Morris:1988cz,sharma2021generalised}. Here, we demonstrate that our modified EWH model, considered in the context of GR, does not evade such a restriction, and exotic matter sources are still required\footnote{Here we do not discuss into detail modified theories of gravity, where traversable wormhole configurations can emerge in the absence of exotic matter sources \cite{Capozziello:2012hr,Guendelman:2013sca,Bambi:2015zch,Ovgun:2018xys,Nascimento:2018sir,Rosa:2018jwp,Fu:2018oaq,Bronnikov:2019ugl,PhysRevD.108.024063}}. For this purpose, it is convenient to introduce the orthonormal frame
\begin{equation}
    \mathbf{e}_{\hat{t}}=\partial_t,\;\mathbf{e}_{\hat{x}}=\partial_x,\;\mathbf{e}_{\hat{\theta}}=\frac{\partial_\theta}{r(x)},\;\mathbf{e}_{\hat{\phi}}=\frac{\partial_\phi}{r(x)\sin\theta},
\end{equation}
that satisfy $g_{\mu\nu}e^{\mu}_{\hat{a}}e^{\nu}_{\hat{b}} = \eta_{\hat{a}\hat{b}}$, where $\eta_{\hat{a}\hat{b}}\equiv \text{diag}(-1,1,1,1)$ is the Minkowski metric. In such a frame the Einstein tensor takes the form $G_{\hat{a}\hat{b}}=G_{\mu\nu}e^{\mu}_{\hat{a}}e^{\nu}_{\hat{b}}$, whose components are:
 \begin{align}
 &G_{\hat{t}\hat{t}} = \dfrac{1-r_{x}^2-2r_{xx}}{r^2},\\  &G_{\hat{x}\hat{x}} = \dfrac{r_{x}^2-1}{r^2},\\
  &G_{\hat{\theta}\hat{\theta}} = G_{\hat{\phi}\hat{\phi}} = \dfrac{r_{xx}}{r}.
 \end{align}
By assuming for the source of the wormhole a fluid description, in the orthonormal basis we can write the energy momentum tensor in the form $T_{\hat{a}\hat{b}}=\text{diag}(\rho(x),-\tau(x),p(x),p(x))$, where its components have a well-known physical interpretation in terms of the energy density $\rho(x)$, the radial tension $\tau(x)$ and the lateral pressure $p(x)$. Now, requiring that our model is a solution of GR, i.e. $G_{\hat{a}\hat{b}}=8\pi T_{\hat{a}\hat{b}}$, we obtain the expressions\footnote{Henceforth we will omit the dependence on $x$ in the components of the energy momentum tensor.}
 \begin{align}
 \rho = \dfrac{1-r_{x}^2-2rr_{xx}}{8\pi r^2},\,\;\tau = \dfrac{1-r_{x}^2}{8\pi r^2},\,\;p = \dfrac{r_{xx}}{8\pi r},
 \end{align}
with $\rho = -2p+\tau$. In GR, energy conditions may be posed in terms of the energy momentum tensor components in the above orthonormal system~\cite{visser2000energy,curiel2017primer}. They are the null energy condition (NEC)
\begin{equation}
 \rho-\tau \geq 0 \text{ and } \rho+p\geq0, 
\end{equation}
the weak energy condition (WEC)
\begin{equation}
 \rho\geq 0,\,\,\rho-\tau \geq 0 \text{ and } \rho+p\geq0, 
\end{equation}
the strong energy condition (SEC)
\begin{equation}
 \rho-\tau+2p\geq 0,\,\,\rho-\tau \geq 0 \text{ and } \rho+p\geq0, 
\end{equation}
and the dominant energy condition (DEC)
\begin{equation}
 \rho\geq 0,\,\,\tau\in[-\rho,\rho] \text{ and } p\in[-\rho,\rho]. 
\end{equation}
As one expects, at the throat all four energy conditions are automatically violated since $r_x(0)=0$ and $r_{xx}(0)>0$, i.e. 
\begin{align}
  \rho\vert_{x=0} = -\dfrac{2ar_{xx}(0)+1}{8\pi a^2}  <0,\\
  (\rho-\tau)\vert_{x=0} = -2p\vert_{x=0} = -\dfrac{2ar_{xx}(0)}{8\pi a^2}  <0.
\end{align}
Since in the inner region the areal radius goes smoothly to a finite value, its first and second derivatives vanishes asymptotically. Therefore, inside the throat, when $r\to R>0$, the lateral pressure vanishes in the very inner region of the object, whereas the density and radial tension are finite and positive in this limit.
The finiteness of these quantities were expected since the curvature scalars are bounded in those models.
On the other hand, when the areal radius shrinks until it vanishes, all the components of $T_{\hat{a}\hat{b}}$ blow up in the very inner region of the object. This is due to the fact that $r$ goes to zero more rapidly than its derivatives, therefore even the lateral pressure diverges in this limit.

\section{Geodesic analysis}\label{sec:geo}
In this section we examine into some details the geodesic trajectories of free point-like particles in their motion over the two regions of the modified EWH. The geodesic equation is obtained from the Lagrangian ${\cal L} = \dot{s}^2/2 = k/2$, where the overdot denotes a derivative with respect to an affine parameter $\lambda$ and $k$ is the normalization of the four-velocity ($k=-1,0$ for massive particles and light-rays, respectively). Due to the symmetries of the Lagrangian, two quantities are conserved along the geodesics, namely $E$ and $\ell$, respectively related with the time translation symmetry (the Lagrangian is independent of $t$) and with the rotational symmetry (the Lagrangian is independent of $\varphi$). Therefore, in the equatorial plane the geodesic equation reads
\begin{equation}
\label{eq:geodesic}
\dot{x}^2=E^2-\left(\dfrac{L^2}{r(x)^2}-k\right),
\end{equation}
and by conducting a thorough analysis of \eqref{eq:geodesic} together with the radial function~\eqref{eq:areal_radius}, we can unveil the underlying geodesic structure of the models we propose.

Let us first consider radial geodesics ($L=0$) moving toward the asymptotic internal region, which, regardless of the asymptotic internal 2-sphere radius $R$, are given in this case by
\begin{align}
 \label{eq:radial_geo_x}   \dot{x}^2&=E^2+k,\\
 \label{eq:radial_geo_r}   \dot{r}^2&=r_x^2(E^2+k).
\end{align}
Upon integration of \eqref{eq:radial_geo_x}, one obtains the trajectory for outgoing particles that cross the throat into the inner region $x(\lambda)= -\sqrt{(E^2+k)}\lambda+x_0$, where $x_0$ is an integration constant. Therefore, one notices that radial geodesics can extend indefinitely, regardless of the asymptotic 2-sphere radius. This is particularly relevant when then inner 2-sphere radius shrinks to zero, and the curvature scalars, energy density and pressures diverge in the asymptotic limit $x\to-\infty$. Since it takes an infinite affine time $\lambda$ to reach the asymptotic infinity, the region with ill-defined quantities is actually inaccessible for massive or massless particles in radial motion. From Eq.~\eqref{eq:radial_geo_r}, one notices that the areal velocity $\dot{r}$ of particles in radial motion goes to zero when $\lambda\to\infty$ (particles going to $x\to-\infty$).

The analysis for non-radial geodesics is more involved, since we cannot obtain an analytical expression for the geodesics. However, some approximations and numerical analysis lead to some interesting conclusions. First, let us consider that the asymptotic radius $R$ is finite and non-vanishing. When it does happen, as one approaches the asymptotic internal region, $\dot{x}^2$ is approximately 
\begin{equation}
\label{eq:geodesic_approx}
\dot{x}^2\approx E^2-\left(\dfrac{L^2}{R^2}-k\right),
\end{equation}
which can also be integrated leading to the conclusion that even non-radial geodesic are complete for both massive and massless particles moving in the internal region of the modified EWH with $R\neq 0$. 

When $R=0$ an interesting feature happens. The effective potential $V_{\text{eff}}=1/r^2(x)$ grows without bound in the internal region, therefore any particle with non-zero angular momentum must suffer a bounce in the internal region being reflected to the outer universe. The only particle capable of propagating indefinitely within this object is one with zero angular momentum, whether it is massive or massless, exhibiting purely radial motion. Hence, even in the vanishing $R$ case, all geodesics are complete.

\subsection{Photon orbits and photon spheres}
To better understand the geodesic structure of the modified EWH, one may study the orbits in these geometries.  
For simplicity let us consider null geodesics ($k=0$), and rewrite Eq.~\eqref{eq:geodesic} as
\begin{equation}
 \label{eq:orbit} \dfrac{1}{r^4}\left(\dfrac{dx}{d\varphi}\right)^2= \dfrac{1}{b^2} - V_{\text{eff}}(x), 
\end{equation}
where $b=L/E$ is the impact parameter that a photon has in the exterior region. Since $x=0$ is a regular minimum of $r^2(x)$, it is also a local maximum of the effective potential. Therefore, the throat radius $a$ plays the role of a critical impact parameter of light-rays, since photons impinging from infinity with impact parameter greater than the throat radius, $b>a$, do not enter in the inner universe and are scattered back to infinity; while photons with impact parameters smaller than the throat radius, $b<a$, do cross the throat and enter in the inner universe; whereas, photons with impact parameter equal to the throat radius, $b=a$, stay trapped in an unstable orbit. The location of this unstable orbit is at the local maximum of the effective potential, that is, $x=0$ or $r=a$. This orbit is called photon sphere, and since it is at an unstable point it is called \textit{unstable} photon sphere. If $r(x)$ has a local maximum, the effective potential also exhibits a local minimum in the inner universe. This local minimum corresponds to a \textit{stable} photon sphere, whose existence may support long-lived modes (radiation may be trapped by these compact object). In Sec.~\ref{sec:SFP} we perform a discussion of trapped scalar modes. 

\begin{figure}
    \centering
    \includegraphics[width=\columnwidth]{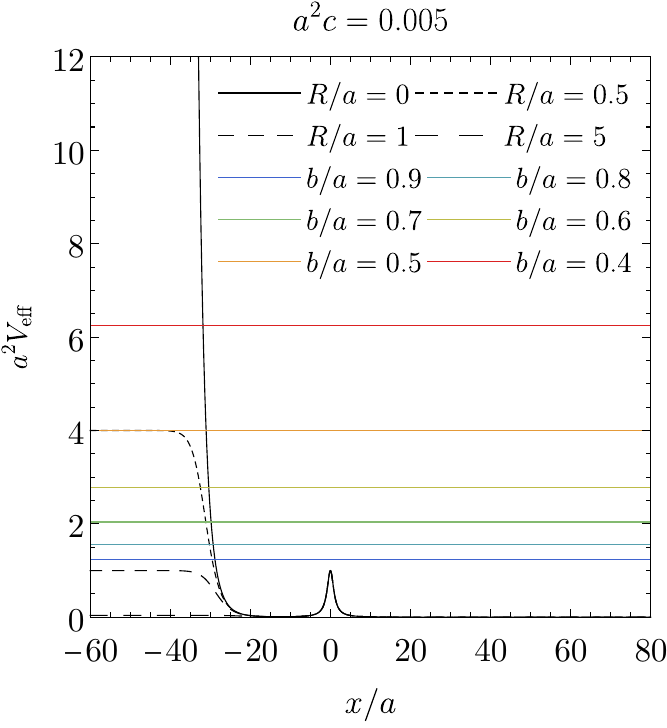}
    \caption{Effective potential of four modified EWHs with fixed $c$ and four choices of $R$, namely $R/a=0,0.5,1$ and 5. We also plot the inverse of the impact parameter squared of some photons that enter in the inner region of the modified EWH.}
    \label{fig:eff_pot}
\end{figure}

The behavior of the effective potential inside the throat deeply departs from the standard profile of the EWH. For a non-vanishing asymptotic 2-sphere radius, the potential goes to a barrier of magnitude $1/R^2$ as $x\to-\infty$, while for $R=0$, the effective potential exponentially grows inside the throat. Any photon able to enter in the inner universe ($b<a$), propagates in a bounded 2-sphere universe, and the behavior of these curves depends on the asymptotic radius $R$. For $R\geq a$, any photon crossing the throat must propagate toward the asymptotic infinity, since the asymptotic value of the effective potential $1/R^2\leq 1/a^2$. 
However for $R<a$, after crossing the throat, depending on the photons' impact parameter, photons may suffer a bounce in the inner universe at $x_b$ and be scattered back to the outer universe, since $1/R^2>1/a^2$. For non-radial geodesics ($L\neq 0$), the bounce happens if given an impact parameter $b$, there is an $x=x_{b}$ such that $V_{\text{eff}}(x_b)=1/b^2$ and $V_{\text{eff}}>1/b^2$ for $x<x_{b}$. At $x_b$ therefore a bounce happens and the particle is scattered to the outer universe. In Fig.~\ref{fig:eff_pot} we show the effective potential of the configurations depicted in Fig.~\ref{fig:embedding_diagrams}, together with the inverse of the impact parameter squared of some photons able to cross the throat ($1/b^2>1/a^2$). The point $x=x_b$ where the vertical lines have the same value of the effective potential is the bounce location, where a photon is reflected in the inner universe and scattered back to the outer one.

\begin{figure*}
    \centering
\includegraphics[width=\columnwidth]{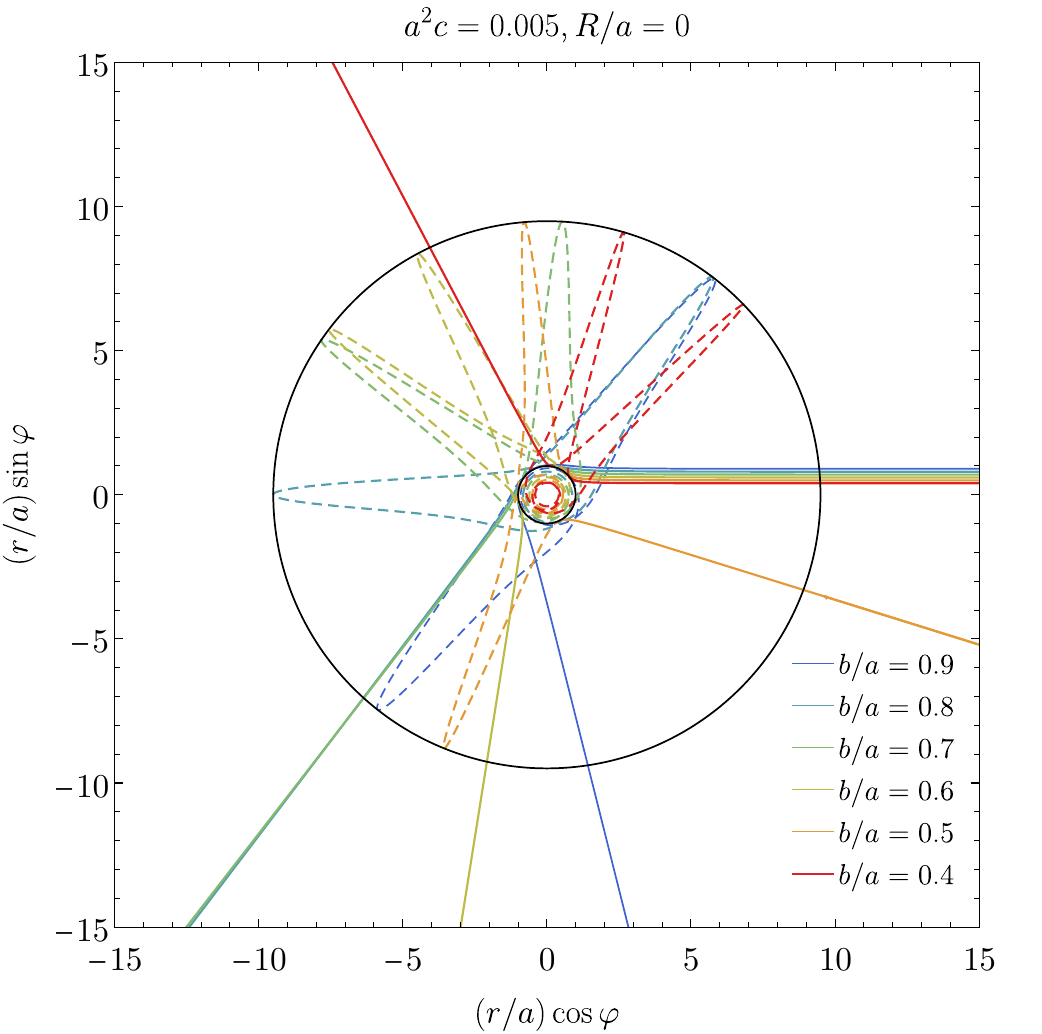}\includegraphics[width=\columnwidth]{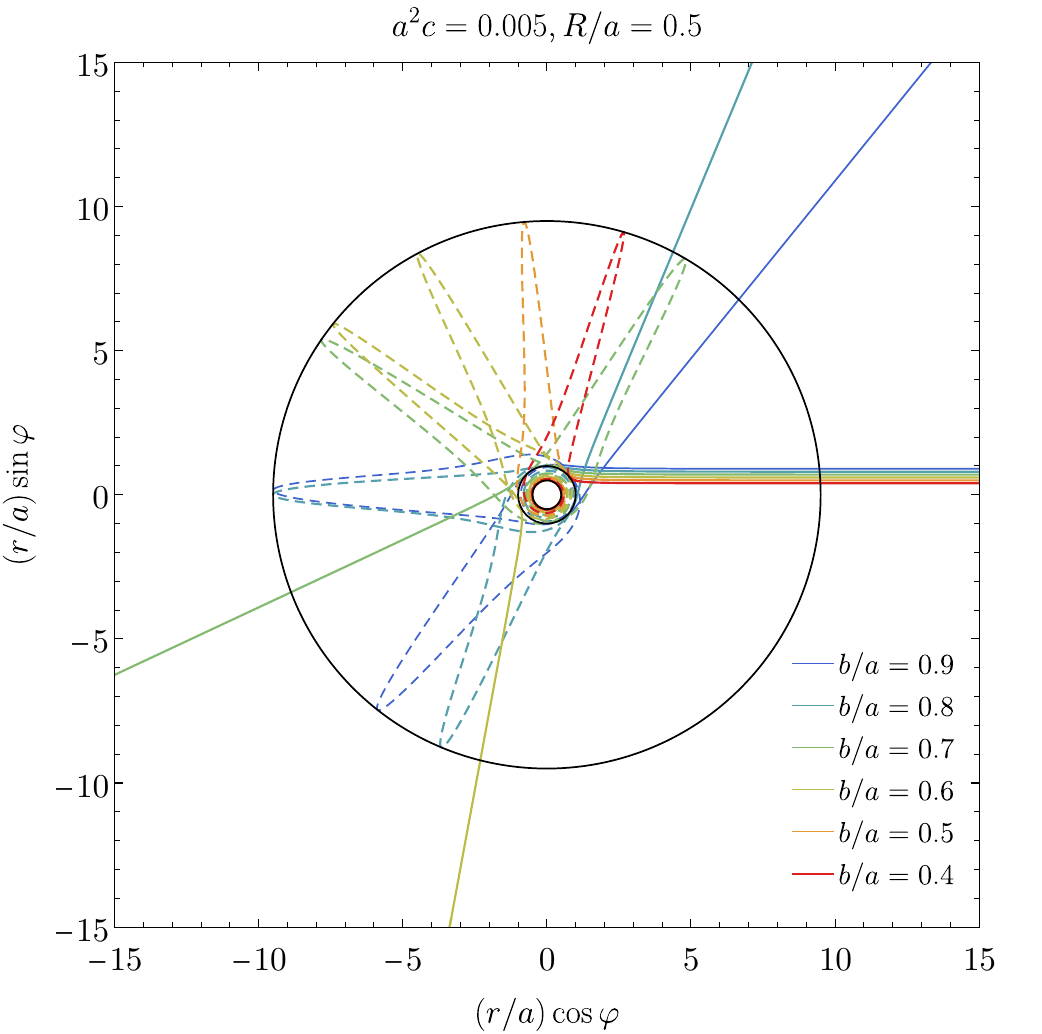}\\
\includegraphics[width=\columnwidth]{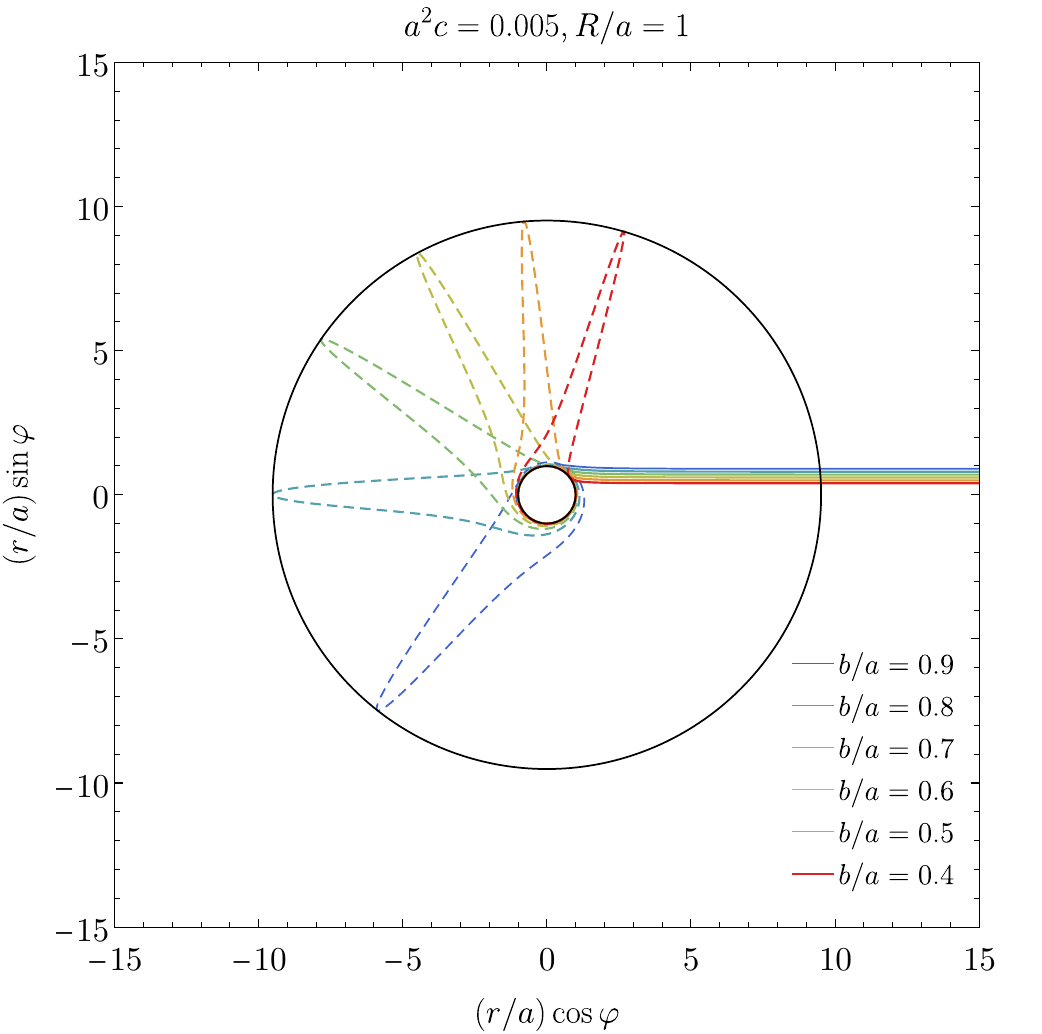}
\includegraphics[width=\columnwidth]{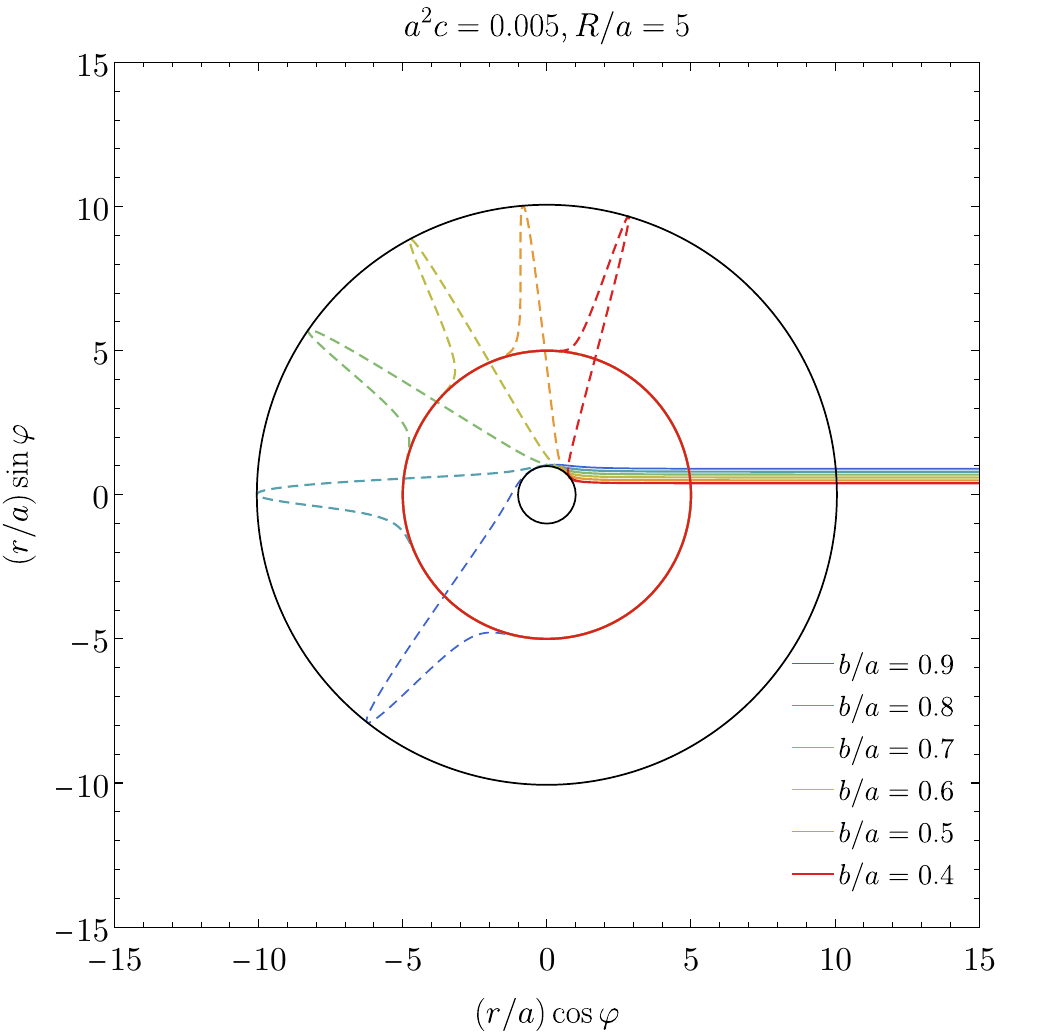}
    \caption{Null geodesics in the modified EWH. Solid lines correspond to light rays propagating in the outer universe, while dashed lines represent geodesics in the inner universe. We are considering photons with the same values of impact parameter in the outer universe, and showing how these photons are scattered or absorbed depending on the modified EWH configuration. The circle with radius 1 corresponds to the throat of the wormhole; the outermost circle corresponds to the local maximum of the areal radius $r(x)$ in the inner universe. The other circles are the asymptotic 2-sphere radius of each configuration.}
    \label{fig:geo_finiteR}
\end{figure*}

In Fig.~\ref{fig:geo_finiteR} we show how the four configurations shown in Fig.~\ref{fig:eff_pot} scatter light rays with the same impact parameter in the outer universe. In the top-left panel, the geodesics propagate in a modified EWH with vanishing asymptotic 2-sphere radius. As previously discussed, any non-radial geodesic moving through the inner region must suffer a bounce. This behavior is shown in the top-left panel of Fig.~\ref{fig:geo_finiteR}. When $R<a$, depending on the impact parameter the geodesic can be scattered to the outer universe or propagate to the asymptotic internal region. We exhibit this behavior in the top-right panel of Fig.~\ref{fig:geo_finiteR}. We can see that, geodesics with impact parameter $b\leq R$ are scattered back to the outer universe, and any other geodesic crossing the throat must propagate to the asymptotic internal region with bounded 2-sphere. The bottom-left and bottom-right panels, respectively, exhibit geodesics in $R=a$ and $R>a$ modified EWHs. In these configurations any photon crossing the throat to the inner universe, propagates toward it to the asymptotic internal region with bounded 2-sphere.

\section{Scalar field perturbations}\label{sec:SFP}

In order to extract valuable information about the geometries investigated in this study, we turn our attention to the analysis of scalar perturbations on the background metric $g_{\mu \nu}$. By examining the evolution of these perturbations, we can discern the distinctive effects that these geometries imprint on the time-domain profiles, making them distinguishable from other black hole and wormhole configurations.

For a massless scalar field $\Psi$ localized within the background metric $g_{\mu \nu}$, the dynamics of the field are governed by the Klein-Gordon equation
\begin{equation}
	\label{eq:KG}
	\Box\Psi(t,x,\theta,\varphi) = 0.
\end{equation}
Due to the spherical symmetry of the problem, we can decompose the field in the following way
\begin{equation}
	\label{eq:field_decom}
	\Psi(x^{\mu}) =\sum_{\ell=0}^{\infty}\sum_{m=-\ell}^{\ell} \frac{\Phi(x,t)}{r(x)}Y_{\ell m}(\theta,\varphi),
\end{equation}
where $Y_{\ell m}(\theta,\varphi)$ denotes spherical harmonics of degree $\ell$ and order $m$. By substituting~\eqref{eq:field_decom} in~\eqref{eq:KG}, one obtains that the radial function 
$\Phi(x,t)$ must satisfy
\begin{align}
	\label{eq:phi}
	\left(\dfrac{d^2}{dt^2}-\dfrac{d^2}{dx^2}+V_{\Phi}\right)\Phi=0,
\end{align}
where the effective potential $V_{\Phi}$ is given by
\begin{equation}
	\label{eq:Vphi} V_{\Phi} = \dfrac{\ell(\ell+1)}{r(x)^2}+\dfrac{r_{xx}}{r(x)}.
\end{equation}

\begin{figure}[b]
	\centering
	\includegraphics[width=0.9\columnwidth]{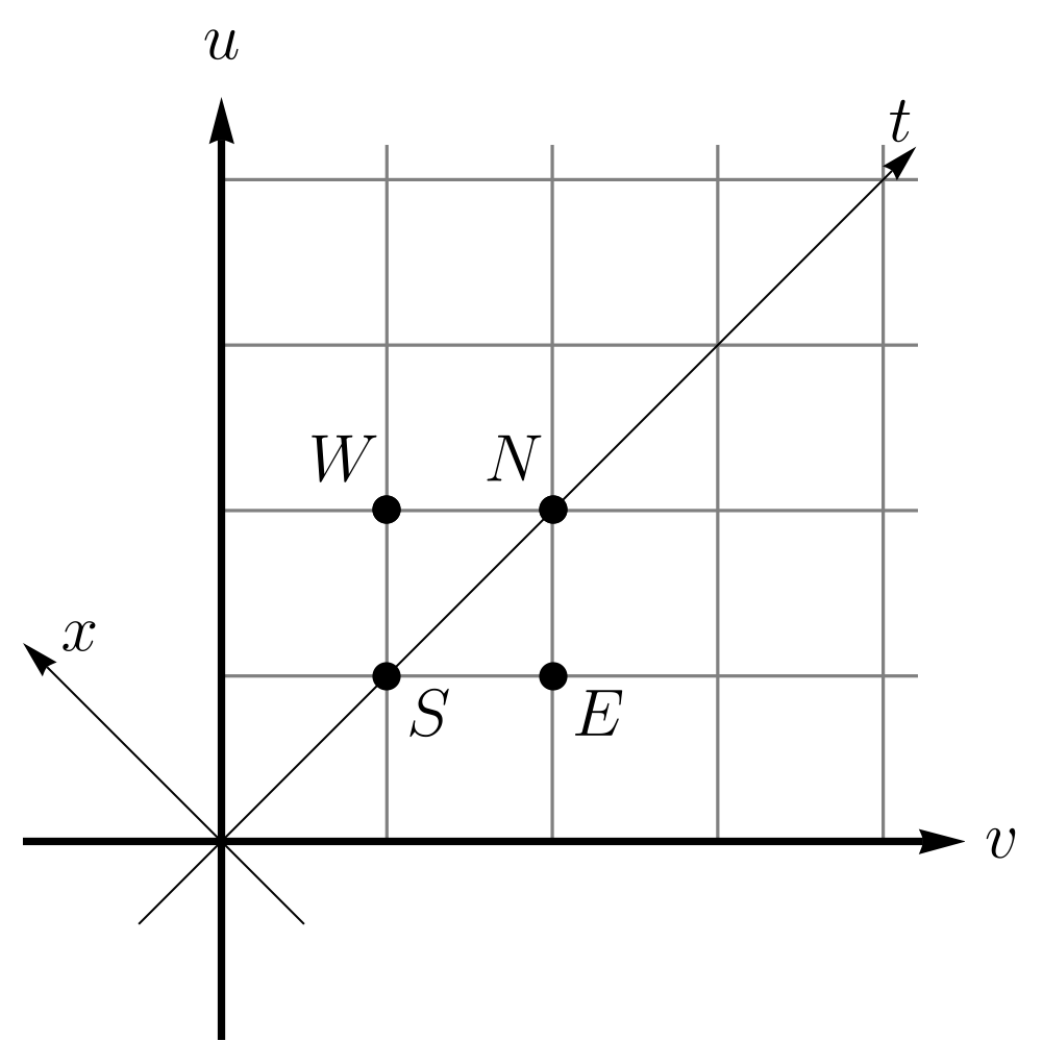}
	\caption{Representation of the numerical grid used for the integration of \eqref{eq:uvdiffeq}. The evaluation points of \eqref{eq:discretizedPhi} are represented with $S$, $W$, $E$ and $N$. The stepsize of the grid can be visualized by the distance between to consecutive points in the same axis $h=u_E-u_S=v_N-v_E$.}
	\label{fig:grid}
\end{figure}

Now, in order to integrate the wave equation \eqref{eq:phi} we follow the procedure described in \cite{Gundlach:1993tp}. This involves introducing light-cone coordinates, specifically the advanced time coordinate denoted as $v\equiv t+x$ and the retarded time coordinate denoted as $u\equiv t-x$. Thus, the wave equation can be expressed as 
\begin{equation}\label{eq:uvdiffeq}
	\left(4\frac{d^2}{dudv}+V_\Phi\right)\Phi=0.
\end{equation}
The integration of this differential equation is done numerically on a null grid which leads to the following expression for the discretized scalar field evolution
\begin{equation}\label{eq:discretizedPhi}
	\Phi_N=\Phi_E+\Phi_W-\Phi_S-\frac{h^2}{8} V_\Phi(S)(\Phi_W+\Phi_E) + O(h^4),
\end{equation}
where $h$ is the stepsize between two neighboring grid points and subscripts indicate the point in the grid where the function is evaluated. Explicitly, $S=(u,v)$, $W=(u+h,v)$, $E=(u,v+h)$ and $N=(u+h,v+h)$, as can be seen more clearly in Fig~\ref{fig:grid}. 

The initial distribution for the scalar perturbation is set on the null surfaces $u=0$ and $v=0$. Then, the grid is computed line by line using the mechanism described in \eqref{eq:discretizedPhi}, with a stepsize $h=0.1$ and grid values ranging from $u_{\text{min}}=0$ to $u_\text{max}=1000$ and $v_{\text{min}}=0$ to $v_\text{max}=1000$. As initial conditions for the scalar perturbation we use a Gaussian distribution on the $u=0$ surface, together with a constant profile on the $v=0$ surface, i.e.:
\begin{equation}
	\Phi(0,v)=Ae^{-(v-v_c)^2/(2\sigma^2)},
\end{equation} 
with height $A=1$, width $\sigma^2=4.5$ and centered at $v_c=20$.

\begin{figure}[t]
	\centering
	\includegraphics[width=\columnwidth]{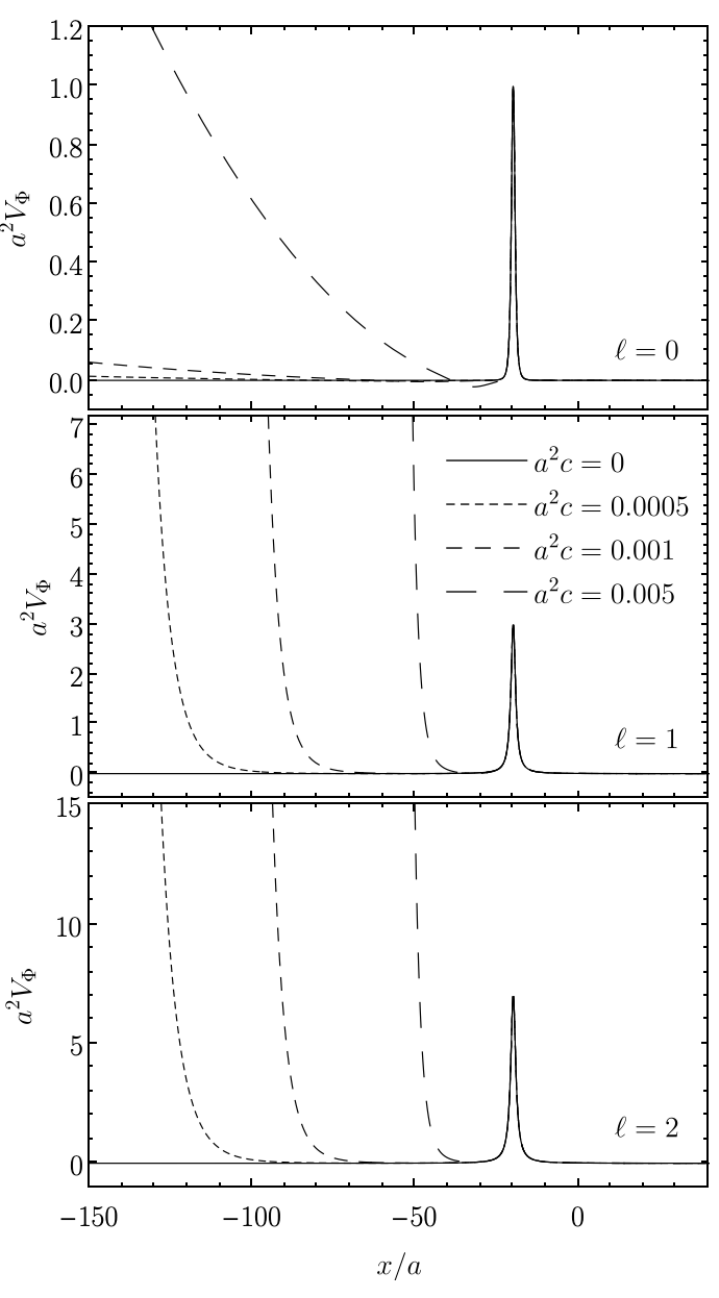}
	\caption{The radial profiles of the effective potential \eqref{eq:Vphi} are presented for three $\ell$-modes and $R/a=0$. The top panel displays the case for $\ell=0$, the central panel shows $\ell=1$, and the bottom panel shows $\ell=2$. The solid line represents the radial profile of the effective potential for the standard EWH, while dashed lines correspond to three different configurations of modified EWH with varying $c$ parameters. }
	\label{fig:SFEP}
\end{figure}

\begin{figure}[h]
	\centering
	\includegraphics[width=\columnwidth]{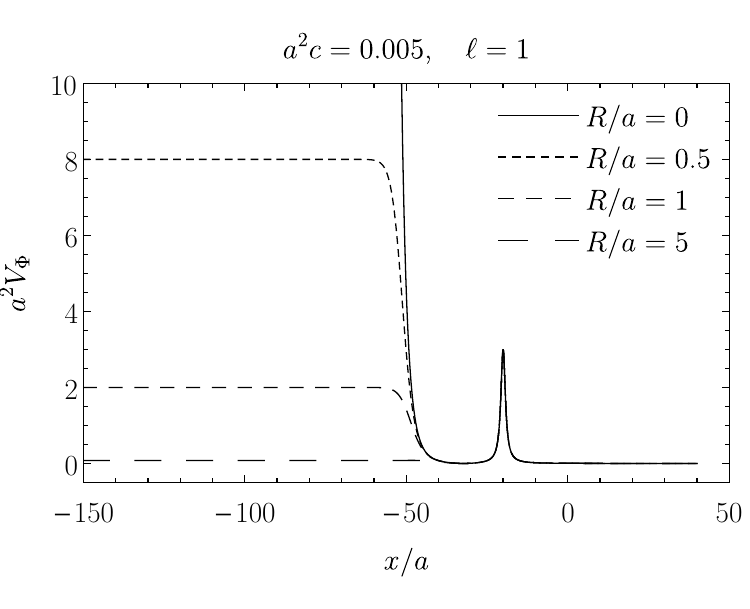}
	\caption{The radial profiles of the effective potential \eqref{eq:Vphi} are presented for four configurations with different $R$ parameter and $\ell=1$. }
	\label{fig:SFEP_R}
\end{figure}

The effective potential is then calculated by applying \eqref{eq:Vphi} and using the radial function $r(x)$ as presented in  \eqref{eq:areal_radius}. Radial profiles of the effective potential with $R/a=0$ are depicted in Fig.~\ref{fig:SFEP}, where, in order to optimize the grid we designed, the throat of the wormhole has been conveniently shifted to $x=-20$. As one can see, the effective potential exhibits a peak like in standard EWHs, which is associated to the throat of the wormhole, and with their maximum value increasing in proportion to $\ell$. Conversely, as $x$ tends to $-\infty$, the potential does not drop to zero, but it shows a rapid and smooth growth, remaining finite for all finite radial values. It is remarkable that the effective potential exhibits a significantly slower growth as $x$ approaches $-\infty$ for the fundamental $\ell$-mode  as compared to higher $\ell$-modes. Moreover, with increasing values of $\ell$, it approaches infinity at a faster rate, although the most prominent discrepancy in growth rate is observed between $\ell=0$ and $\ell=1$. Such a behavior, combined with the first peak, gives rise to a well, which is expected to lead to echoes in the time-domain spectrum (see discussion in the next section). For the fundamental $\ell$-mode, the effective potential assumes negative values within the well near the peak associated with the throat of the wormhole. The radial extent over which negative effective potential occurs diminishes as $\ell$ increases. Finally, it is worth mentioning that the parameter $c$ exerts an influence on the effective potential, causing the well formed between the throat peak and the asymptotic boundary to narrow as its value increases.

For the non-vanishing $R$ case, one can show that the asymptotic behavior of $V_{\Phi}$ reads
\begin{align}
	\label{eq:V_xinf}
	&\lim_{x\to +\infty}V_{\Phi} = 0,\\
	\label{eq:V_xminf} 
	&\lim_{x\to-\infty}V_{\Phi} = \left\{\begin{array}{ll}    0, &\ell=0,\\
 \dfrac{\ell(\ell+1)}{R^2}, &\ell\neq 0.
 \end{array}\right.
\end{align} 
 It is worth nothing that in the asymptotic internal region, whenever $\ell\neq 0$ and $R\neq 0$, the effective potential goes to a threshold value. The massless scalar field propagating in this region would behave equivalently to a scalar field with effective mass in a Minkowski background, namely 
\begin{equation}
    \label{eq:we_asymptoticregion2}
    \left(\dfrac{d^2}{dt^2}-\dfrac{d^2}{dx^2}+\mu_e^2\right)\Phi=0,
\end{equation}
where $\mu_e=\sqrt{\ell(\ell+1)}/R$. This kind of effective mass in scalar field dynamics typically arises in non-asymptotically flat spacetimes, such as when scalar waves propagate around a black hole immersed in a magnetic field~\cite{kokkotas2011quasinormal,turimov2019quasinormal}. For the case $\ell=0$, the effective potential vanishes asymptotically in both sides of the modified EWH, therefore no effective mass term appears.

In Fig.~\ref{fig:SFEP_R}, we depict the radial profiles of the effective potential for $\ell=1$ and various values of the $R$ parameter. As observed in the geodesic analysis, the presence of a non-zero $R$ leads to the asymptotic finiteness of the potential as $x\to-\infty$. The finite value towards which the potential tends is inversely proportional to the magnitude of $R$, with the potential becoming approximately two orders of magnitude smaller than the throat peak for $R/a=5$. 

As the 2-sphere approaches its asymptotic value, the effective potential exhibits a sort of effective centrifugal barrier for $R=0$, playing the role of an effective mirror for the scalar field perturbation. For $R\neq 0$ in this same limit, one can see that the effective potential tends to a plateau which height is proportional to $1/R^2$.


\subsection{Discussion of the results}

\begin{figure*}
	\centering
	\includegraphics[width=\columnwidth]{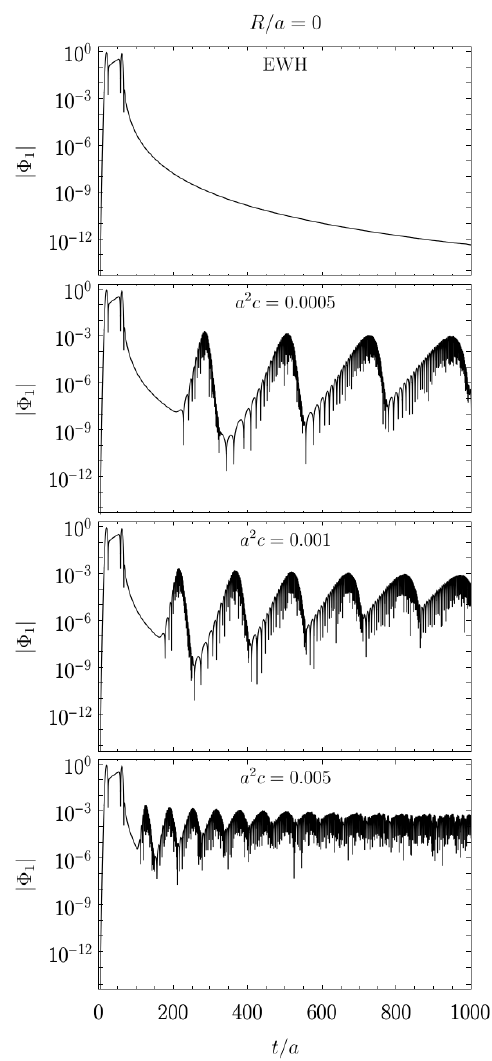}
	\includegraphics[width=\columnwidth]{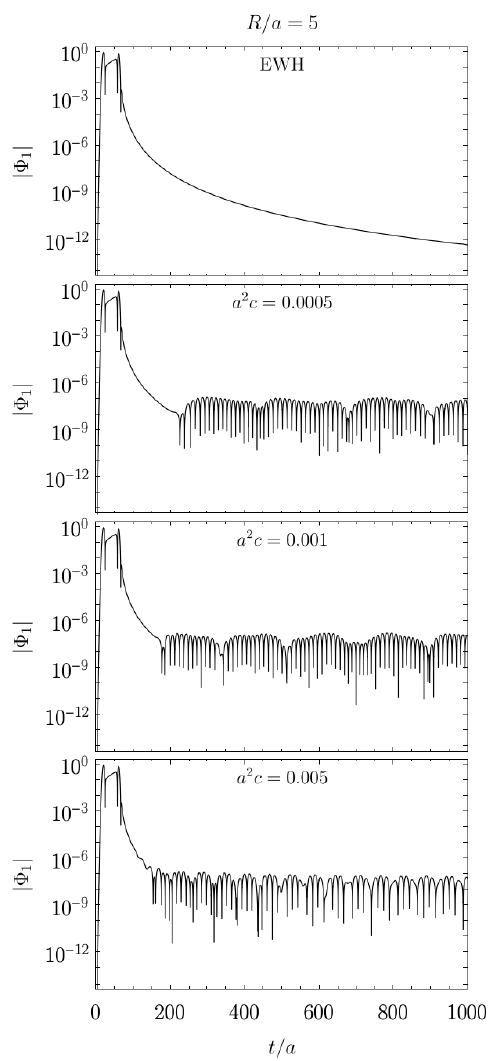}
	\caption{Time-domain profile of the absolute value of the scalar field perturbation for the $\ell=1$ mode. The top panels depict identical geometries associated with the EWH. In the left column, three configurations with varying values of the parameter $c$ and $R/a=0$ are displayed. In the right column, three configurations with different values of the parameter $c$ are shown, but with $R/a=5$. The disparity between the EWH and the other cases are the so-called echoes.  }
	\label{fig:echoes_R0andR5}
\end{figure*}

After solving the discretized wave equation, we extract the scalar field values at the observation point $x_\text{obs}=0$ using the coordinate transformation $x=(u-v)/2$ and $t=(u+v)/2$. The numerical integration results are displayed in Fig.~\ref{fig:echoes_R0andR5}. In the top panels, the time-domain evolution of the scalar field for an EWH is shown as a reference. Even though our model allows arbitrary large values of $c$, here we are focusing on small deviations of EWHs in the throat scale, therefore restricting our analysis to $a^2c\leq 0.01$. This makes with the potential peak at the throat to be almost the same as EWH (see Figs.~\ref{fig:SFEP} and~\ref{fig:SFEP_R}). As the initial wave packet impinges on the modified EWH, it encounters the throat peak first, causing a portion of the wave to return to the observation point and exhibiting a characteristic ringdown. Due to the similarities between the throat peaks of modified and standard EWHs, the prompt contribution and initial ringdown signal in the time profile of scalar perturbations are expected to be basically the same in both scenarios.


The portion of the scalar wave packet that is transmitted by the throat peak passes through the well and encounters the potential barrier extending to $x\to-\infty$. It then reflects back towards the observation point. However, in order to reach $x_\text{obs}$, it must interact with the throat peak once again. A portion of the incident wave is reflected, repeating the same process, while another portion is transmitted. The transmitted portion, after being detected at the observation point, evolves towards $x\to\infty$ and does not pass through the observation point again. Each time a wave is reflected and passes through the observation point, it is recorded as a peak in the time-domain profile of $|\Phi_l|$. As expected based on the description of the effective potential of modified EWH, (unstable) photon sphere modes from the scalar perturbation exist and ring in the same way as in standard EWH. However, in this case, there is a stable photon sphere in the inner universe, causing the bounded 2-sphere region of the wormhole to act as a cavity, trapping photon sphere modes. This results in a series of echoes in the scalar perturbation time profile, which are illustrated in Fig.~\ref{fig:echoes_R0andR5}. The characteristics of these echoes vary depending on the parameters used to construct the spacetime, since the width and height of the cavity in the effective potential deeply depend on $c$ and $R$.

Let us now discuss the $R=0$ spacetimes, corresponding to the left column of Fig.~\ref{fig:echoes_R0andR5}. It can be observed that there is a relationship between the value of the parameter $c$ and the frequency of the echoes. This arises from the fact that $c$ influences the width of the effective potential well. As the well becomes narrower with increasing $c$, it can be observed that the time interval between two consecutive echoes decreases (or, equivalently the frequency increases), as the waves have to travel a shorter distance.

Another notable feature that can be observed is the gradual decay of peak amplitudes. Each time the scalar wave packet interacts with the throat peak, it is divided into transmitted and reflected parts, resulting in weaker successive echoes. This effect contrasts with the case of the EWH, where the signal exhibits a rapid decay compared to the modified Ellis scenario. The presence of two asymptotic infinities and the absence of a potential well in the EWH prevent the emergence of echoes, thereby contributing to the faster signal decay.
When the potential well is narrower, the superposition of different echoes becomes more noticeable, leading to deformations in their waveforms. This behavior is particularly evident in the late-time regime, as shown by the bottom plot in the left column of Fig.~\ref{fig:echoes_R0andR5}.

Similar features are noticeable also in the right column of Fig.~\ref{fig:echoes_R0andR5}, corresponding to the $R/a=5$ configuration. However, the echoes' amplitudes are notably smaller compared to the previous scenario, making it difficult to differentiate distinct echoes due to their reduced amplitude.

\begin{figure}
	\centering
	\includegraphics[width=\columnwidth]{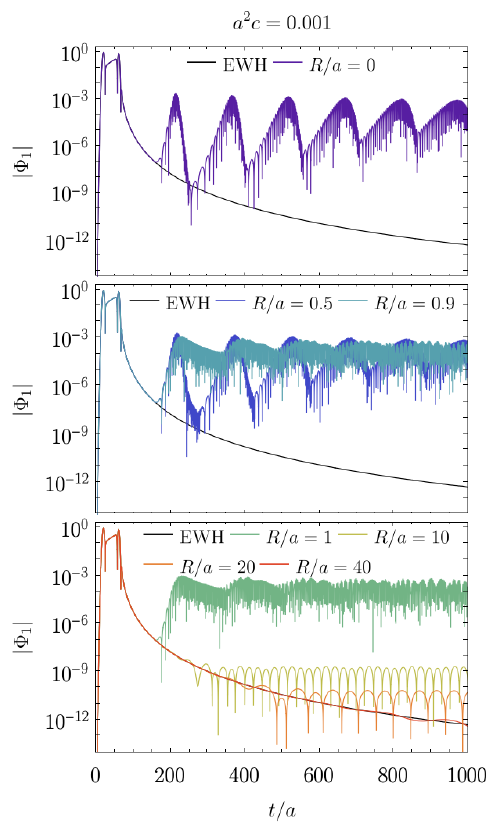}
	\caption{Time-domain profile of the absolute value of the scalar field perturbation for the $\ell=1$ mode. Seven configurations with the following values $R/a=\left\{0,0.5,0.9,1,10,20,40\right\}$ are plotted, along with the EWH as a reference.}
	\label{fig:echoes_Rdependent}
\end{figure}

The absence of an unbounded growing effective potential results in wave packets interacting with a finite barrier, leading once again to the division of the package into a reflected part and a transmitted part that propagates towards $x\to-\infty$ without bouncing back to the outer universe. The height of the barrier diminishes as $R$ increases, making it evident that the reflected part of the wave also is smaller as one considers bigger values of $R$, as shown in Fig.~\ref{fig:echoes_Rdependent}. Notably, there is a distinct transition regime depending on the asymptotic value to which the effective potential tends. Specifically, the asymptotic value aligns with the height of the throat peak for $R/a \approx 0.8$ for the considered configurations. When the barrier exceeds the throat peak height ($R/a < 0.8$), the echoes are easily distinguishable. However, when the barrier is smaller ($R/a>0.8$), a higher proportion of the wave is lost, leading to a reduction in the amplitude of the echoes.

For late-times, we observe a diminishment in the damping of the echoes of the scalar wave. At this point, so-called quasi-resonances are present---arbitrarily long-lived QNMs~\cite{ohashi2004massive}.
These modes were discussed in the QNM analysis of massive fields in asymptotically flat spacetimes~\cite{ohashi2004massive,konoplya2005decay}, and also appears in non-asymptotically flat scenarios where the wave equation acquires a sort of effective mass~\cite{turimov2019quasinormal,shao2022quasinormal}. Our model pertains to the latter case.

Remarkably, in the asymptotic limit $R \rightarrow \infty$, the effective potential $V_{\Phi}$ recovers the usual behaviour of the one of EWH in the inner universe, that is, $V_{\Phi}\to 0$ as $x\to-\infty$. Therefore, the time-domain profile of modified EWH with large asymptotic 2-sphere radius, tends towards the expected profile of a standard EWH.

Finally, it is noteworthy that while the results presented here are for the $\ell=1$ mode, the qualitative characteristics are present in the other $\ell$-modes as well. 

\section{Final remarks}\label{sec:conclusion}
We have introduced a new class of modified EWHs, where one side of the global spacetime geometry, usually consisting in an asymptotic Minkowski region, is supplanted by a bounded 2-sphere patch, where the areal radius is finite and asymptotically constant. The resulting asymmetric wormhole is still endowed with a local minimum in the areal radius, i.e. a throat-like structure, where a transition between the two regions takes place. This was achieved by means of a modified radial function, which smoothly interpolates from the standard Ellis case to the bounded 2-sphere core on the other side. 

In the internal region of the modified EWH, the areal radius exhibits a local maximum and eventually collapses to its asymptotic value $R$. Such a peculiar behavior is ultimately responsible for a rich geodesic structure. We performed an extensive analysis on causal geodesics, finding that regardless of the model parameters, the modified EWH is geodesically complete, even in the vanishing asymptotic 2-sphere scenario, $R=0$. This is indeed relevant, because in the vanishing 2-sphere radius case, 
the curvature scalars, energy density and pressures are unbounded as $x\to-\infty$, but these divergences remain unreachable for causal geodesics. In the $R\neq 0$ scenario, the curvature scalars are well-defined everywhere and go to the ones of a 2-sphere with radius $R$ in the asymptotic internal region.

We also investigated photon orbits in this model by numerically solving the geodesic equation. When the asymptotic value of the 2-spheres radius is lower than the throat radius, $R<a$, there are light rays that depending on their impact parameter suffer a bounce in the inner universe and are scattered back to the outer universe. As $R$ diminishes, and the inner region becomes more bounded, the photon's impact parameter necessary for the bounce to happen is smaller (see Fig.~\ref{fig:eff_pot}). The extreme scenario is then for $R=0$, where all geodesics entering the inner universe are scattered back to the outer one, and the only null geodesic that can propagate forever in the inner region is the one of a photon in purely radial motion.
Lastly, for $R\geq a$, any photon that crosses the throat propagates towards the asymptotic infinity with bounded 2-sphere radius.

The analysis of the effective potential shows that an unstable photon sphere is present at the throat, similarly to the standard EWH. Moreover, the inner universe has an additional structure, a stable photon sphere, that is located in the local minimum of the effective potential. The existence of the stable photon sphere allows long-lived modes to be trapped, giving rise to the appearance of echoes in the ringdown profile, usually absent in standard EWHs. In order to probe this aspect, we performed an analysis of massless scalar excitations in the modified EWH background. 
After being transmitted to the inner universe the scalar perturbation interacts with an effective centrifugal barrier for $R=0$ or a step potential for $R\neq 0$. Therefore, part of the modes are trapped in a potential well and observed as a series of echoes in the time profile of the scalar perturbation. In particular, we noticed that the role of the parameter $c$ is to modify the time delay between two successive echoes. When $c$ increases, the cavity in the effective potential becomes narrower and the echoes' time delay diminishes. 

The role of the asymptotically bounded 2-sphere radius $R$ is more subtle. For the $R=0$ case, the scalar perturbations get trapped between the unbounded barrier and the photon sphere at the throat. By increasing $R$, this behavior persists, but since the unbounded barrier is replaced by a step potential barrier, part of the modes can be transmitted through the asymptotic internal region.

We notice that by replacing the asymptotically Minkowski spacetime with a bounded 2-sphere core, the scalar field acquires an effective mass in the asymptotic internal region for non-vanishing values of the multipole number $\ell$ and non-zero asymptotic 2-sphere radius $R$. A similar effect is observed with scalar waves propagating around magnetized black holes~\cite{turimov2019quasinormal}. This leads to the presence of quasi-resonances in the late-time profile, which are arbitrarily long-lived QNMs. Whether $R=0$ and $\ell\neq 0$, the effective mass grows without bound acting as an effective mirror for the scalar wave. Consequently, the scalar perturbation cannot reach the asymptotic internal region, and the modes trapped near the stable photon sphere at some late time should tunnel to the outer universe.

Our results indicate that the observational features of wormholes are crucially dependent on their global characteristics, even though they can share a very similar throat-like structure. The study of perturbations can be used to extract valuable information about the  compactness of the inner universe, and this could be useful in future spectroscopy experiments trying to identify new compact objects.
We remark that the geometrical structure of the model described in the present work can be enriched further either by considering Schwarzschild-like scenarios, or by restricting the domain of the bounded 2-sphere patch, i.e. truncating the range of the possible values spanned by the coordinate $x$ and imposing specific boundary conditions. These configurations are expected to generate a great variety of phenomenological signatures due to the appearance of additional potential barriers or closed universe effects in geodesic motion.  The study of such aspects is currently underway and will be the subject of forthcoming works.
\begin{acknowledgments}
		The authors would like to acknowledge 
		Funda\c{c}\~ao Amaz\^onia de Amparo a Estudos e Pesquisas (FAPESPA), 
		Conselho Nacional de Desenvolvimento Cient\'ifico e Tecnol\'ogico (CNPq)
		and Coordena\c{c}\~ao de Aperfei\c{c}oamento de Pessoal de N\'ivel Superior (CAPES) -- Finance Code 001, from Brazil, for partial financial support. 
  This work is supported by the Spanish Grant PID2020-116567GB- C21 funded by MCIN/AEI/10.13039/501100011033, and the project PROMETEO/2020/079 (Generalitat Valenciana). This research has further been supported by the European Union's Horizon 2020 research and innovation (RISE) programme H2020-MSCA-RISE-2017 Grant No. FunFiCO-777740 and by the European Horizon Europe staff exchange (SE) programme HORIZON-MSCA-2021-SE-01 Grant No. NewFunFiCO-101086251. 
  LC thanks University of Aveiro, in Portugal, for the kind hospitality during the completion of this work.
  AMF is supported by the Spanish Ministerio de Cienciea y Innovación with the PhD fellowship PRE2018-083802. The work of FB is supported by the postdoctoral grant CIAPOS/2021/169. 
	\end{acknowledgments}

\bibliography{bibliography}

\begin{thebibliography}{52}%
\makeatletter
\providecommand \@ifxundefined [1]{%
 \@ifx{#1\undefined}
}%
\providecommand \@ifnum [1]{%
 \ifnum #1\expandafter \@firstoftwo
 \else \expandafter \@secondoftwo
 \fi
}%
\providecommand \@ifx [1]{%
 \ifx #1\expandafter \@firstoftwo
 \else \expandafter \@secondoftwo
 \fi
}%
\providecommand \natexlab [1]{#1}%
\providecommand \enquote  [1]{``#1''}%
\providecommand \bibnamefont  [1]{#1}%
\providecommand \bibfnamefont [1]{#1}%
\providecommand \citenamefont [1]{#1}%
\providecommand \href@noop [0]{\@secondoftwo}%
\providecommand \href [0]{\begingroup \@sanitize@url \@href}%
\providecommand \@href[1]{\@@startlink{#1}\@@href}%
\providecommand \@@href[1]{\endgroup#1\@@endlink}%
\providecommand \@sanitize@url [0]{\catcode `\\12\catcode `\$12\catcode
  `\&12\catcode `\#12\catcode `\^12\catcode `\_12\catcode `\%12\relax}%
\providecommand \@@startlink[1]{}%
\providecommand \@@endlink[0]{}%
\providecommand \url  [0]{\begingroup\@sanitize@url \@url }%
\providecommand \@url [1]{\endgroup\@href {#1}{\urlprefix }}%
\providecommand \urlprefix  [0]{URL }%
\providecommand \Eprint [0]{\href }%
\providecommand \doibase [0]{https://doi.org/}%
\providecommand \selectlanguage [0]{\@gobble}%
\providecommand \bibinfo  [0]{\@secondoftwo}%
\providecommand \bibfield  [0]{\@secondoftwo}%
\providecommand \translation [1]{[#1]}%
\providecommand \BibitemOpen [0]{}%
\providecommand \bibitemStop [0]{}%
\providecommand \bibitemNoStop [0]{.\EOS\space}%
\providecommand \EOS [0]{\spacefactor3000\relax}%
\providecommand \BibitemShut  [1]{\csname bibitem#1\endcsname}%
\let\auto@bib@innerbib\@empty
\bibitem [{\citenamefont {Abbott}\ \emph {et~al.}(2016)\citenamefont {Abbott}
  \emph {et~al.}}]{LIGOScientific:2016aoc}%
  \BibitemOpen
  \bibfield  {author} {\bibinfo {author} {\bibfnamefont {B.~P.}\ \bibnamefont
  {Abbott}} \emph {et~al.} (\bibinfo {collaboration} {LIGO Scientific,
  Virgo}),\ }\href {https://doi.org/10.1103/PhysRevLett.116.061102} {\bibfield
  {journal} {\bibinfo  {journal} {Phys. Rev. Lett.}\ }\textbf {\bibinfo
  {volume} {116}},\ \bibinfo {pages} {061102} (\bibinfo {year} {2016})},\
  \Eprint {https://arxiv.org/abs/1602.03837} {arXiv:1602.03837 [gr-qc]}
  \BibitemShut {NoStop}%
\bibitem [{\citenamefont {Akiyama}\ \emph {et~al.}(2019)\citenamefont {Akiyama}
  \emph {et~al.}}]{EventHorizonTelescope:2019dse}%
  \BibitemOpen
  \bibfield  {author} {\bibinfo {author} {\bibfnamefont {K.}~\bibnamefont
  {Akiyama}} \emph {et~al.} (\bibinfo {collaboration} {Event Horizon
  Telescope}),\ }\href {https://doi.org/10.3847/2041-8213/ab0ec7} {\bibfield
  {journal} {\bibinfo  {journal} {Astrophys. J. Lett.}\ }\textbf {\bibinfo
  {volume} {875}},\ \bibinfo {pages} {L1} (\bibinfo {year} {2019})},\ \Eprint
  {https://arxiv.org/abs/1906.11238} {arXiv:1906.11238 [astro-ph.GA]}
  \BibitemShut {NoStop}%
\bibitem [{\citenamefont {Herdeiro}\ and\ \citenamefont
  {Radu}(2014)}]{Herdeiro:2014goa}%
  \BibitemOpen
  \bibfield  {author} {\bibinfo {author} {\bibfnamefont {C.~A.~R.}\
  \bibnamefont {Herdeiro}}\ and\ \bibinfo {author} {\bibfnamefont
  {E.}~\bibnamefont {Radu}},\ }\href
  {https://doi.org/10.1103/PhysRevLett.112.221101} {\bibfield  {journal}
  {\bibinfo  {journal} {Phys. Rev. Lett.}\ }\textbf {\bibinfo {volume} {112}},\
  \bibinfo {pages} {221101} (\bibinfo {year} {2014})},\ \Eprint
  {https://arxiv.org/abs/1403.2757} {arXiv:1403.2757 [gr-qc]} \BibitemShut
  {NoStop}%
\bibitem [{\citenamefont {Herdeiro}\ \emph {et~al.}(2016)\citenamefont
  {Herdeiro}, \citenamefont {Radu},\ and\ \citenamefont
  {R\'unarsson}}]{Herdeiro:2016tmi}%
  \BibitemOpen
  \bibfield  {author} {\bibinfo {author} {\bibfnamefont {C.}~\bibnamefont
  {Herdeiro}}, \bibinfo {author} {\bibfnamefont {E.}~\bibnamefont {Radu}},\
  and\ \bibinfo {author} {\bibfnamefont {H.}~\bibnamefont {R\'unarsson}},\
  }\href {https://doi.org/10.1088/0264-9381/33/15/154001} {\bibfield  {journal}
  {\bibinfo  {journal} {Class. Quant. Grav.}\ }\textbf {\bibinfo {volume}
  {33}},\ \bibinfo {pages} {154001} (\bibinfo {year} {2016})},\ \Eprint
  {https://arxiv.org/abs/1603.02687} {arXiv:1603.02687 [gr-qc]} \BibitemShut
  {NoStop}%
\bibitem [{\citenamefont {Dymnikova}(1992)}]{Dymnikova:1992ux}%
  \BibitemOpen
  \bibfield  {author} {\bibinfo {author} {\bibfnamefont {I.}~\bibnamefont
  {Dymnikova}},\ }\href {https://doi.org/10.1007/BF00760226} {\bibfield
  {journal} {\bibinfo  {journal} {Gen. Rel. Grav.}\ }\textbf {\bibinfo {volume}
  {24}},\ \bibinfo {pages} {235} (\bibinfo {year} {1992})}\BibitemShut
  {NoStop}%
\bibitem [{\citenamefont {Ansoldi}(2008)}]{Ansoldi:2008jw}%
  \BibitemOpen
  \bibfield  {author} {\bibinfo {author} {\bibfnamefont {S.}~\bibnamefont
  {Ansoldi}},\ }in\ \href@noop {} {\emph {\bibinfo {booktitle} {{Conference on
  Black Holes and Naked Singularities}}}}\ (\bibinfo {year} {2008})\ \Eprint
  {https://arxiv.org/abs/0802.0330} {arXiv:0802.0330 [gr-qc]} \BibitemShut
  {NoStop}%
\bibitem [{\citenamefont {Liebling}\ and\ \citenamefont
  {Palenzuela}(2023)}]{Liebling:2012fv}%
  \BibitemOpen
  \bibfield  {author} {\bibinfo {author} {\bibfnamefont {S.~L.}\ \bibnamefont
  {Liebling}}\ and\ \bibinfo {author} {\bibfnamefont {C.}~\bibnamefont
  {Palenzuela}},\ }\href {https://doi.org/10.1007/s41114-023-00043-4}
  {\bibfield  {journal} {\bibinfo  {journal} {Living Rev. Rel.}\ }\textbf
  {\bibinfo {volume} {26}},\ \bibinfo {pages} {1} (\bibinfo {year} {2023})},\
  \Eprint {https://arxiv.org/abs/1202.5809} {arXiv:1202.5809 [gr-qc]}
  \BibitemShut {NoStop}%
\bibitem [{\citenamefont {Visinelli}(2021)}]{Visinelli:2021uve}%
  \BibitemOpen
  \bibfield  {author} {\bibinfo {author} {\bibfnamefont {L.}~\bibnamefont
  {Visinelli}},\ }\href {https://doi.org/10.1142/S0218271821300068} {\bibfield
  {journal} {\bibinfo  {journal} {Int. J. Mod. Phys. D}\ }\textbf {\bibinfo
  {volume} {30}},\ \bibinfo {pages} {2130006} (\bibinfo {year} {2021})},\
  \Eprint {https://arxiv.org/abs/2109.05481} {arXiv:2109.05481 [gr-qc]}
  \BibitemShut {NoStop}%
\bibitem [{\citenamefont {Mazur}\ and\ \citenamefont
  {Mottola}(2023)}]{mazur2023gravitational}%
  \BibitemOpen
  \bibfield  {author} {\bibinfo {author} {\bibfnamefont {P.~O.}\ \bibnamefont
  {Mazur}}\ and\ \bibinfo {author} {\bibfnamefont {E.}~\bibnamefont
  {Mottola}},\ }\href@noop {} {\bibfield  {journal} {\bibinfo  {journal}
  {Universe}\ }\textbf {\bibinfo {volume} {9}},\ \bibinfo {pages} {88}
  (\bibinfo {year} {2023})}\BibitemShut {NoStop}%
\bibitem [{\citenamefont {Visser}(1995)}]{Visser:1995cc}%
  \BibitemOpen
  \bibfield  {author} {\bibinfo {author} {\bibfnamefont {M.}~\bibnamefont
  {Visser}},\ }\href@noop {} {\emph {\bibinfo {title} {{Lorentzian wormholes:
  From Einstein to Hawking}}}}\ (\bibinfo {year} {1995})\BibitemShut {NoStop}%
\bibitem [{\citenamefont {Simpson}\ and\ \citenamefont
  {Visser}(2019)}]{Simpson:2018tsi}%
  \BibitemOpen
  \bibfield  {author} {\bibinfo {author} {\bibfnamefont {A.}~\bibnamefont
  {Simpson}}\ and\ \bibinfo {author} {\bibfnamefont {M.}~\bibnamefont
  {Visser}},\ }\href {https://doi.org/10.1088/1475-7516/2019/02/042} {\bibfield
   {journal} {\bibinfo  {journal} {JCAP}\ }\textbf {\bibinfo {volume} {02}},\
  \bibinfo {pages} {042}},\ \Eprint {https://arxiv.org/abs/1812.07114}
  {arXiv:1812.07114 [gr-qc]} \BibitemShut {NoStop}%
\bibitem [{\citenamefont {Morris}\ and\ \citenamefont
  {Thorne}(1988)}]{Morris:1988cz}%
  \BibitemOpen
  \bibfield  {author} {\bibinfo {author} {\bibfnamefont {M.~S.}\ \bibnamefont
  {Morris}}\ and\ \bibinfo {author} {\bibfnamefont {K.~S.}\ \bibnamefont
  {Thorne}},\ }\href {https://doi.org/10.1119/1.15620} {\bibfield  {journal}
  {\bibinfo  {journal} {Am. J. Phys.}\ }\textbf {\bibinfo {volume} {56}},\
  \bibinfo {pages} {395} (\bibinfo {year} {1988})}\BibitemShut {NoStop}%
\bibitem [{\citenamefont {Maldacena}\ and\ \citenamefont
  {Susskind}(2013)}]{Maldacena:2013xja}%
  \BibitemOpen
  \bibfield  {author} {\bibinfo {author} {\bibfnamefont {J.}~\bibnamefont
  {Maldacena}}\ and\ \bibinfo {author} {\bibfnamefont {L.}~\bibnamefont
  {Susskind}},\ }\href {https://doi.org/10.1002/prop.201300020} {\bibfield
  {journal} {\bibinfo  {journal} {Fortsch. Phys.}\ }\textbf {\bibinfo {volume}
  {61}},\ \bibinfo {pages} {781} (\bibinfo {year} {2013})},\ \Eprint
  {https://arxiv.org/abs/1306.0533} {arXiv:1306.0533 [hep-th]} \BibitemShut
  {NoStop}%
\bibitem [{\citenamefont {Nandi}\ \emph {et~al.}(2006)\citenamefont {Nandi},
  \citenamefont {Zhang},\ and\ \citenamefont {Zakharov}}]{Nandi:2006ds}%
  \BibitemOpen
  \bibfield  {author} {\bibinfo {author} {\bibfnamefont {K.~K.}\ \bibnamefont
  {Nandi}}, \bibinfo {author} {\bibfnamefont {Y.-Z.}\ \bibnamefont {Zhang}},\
  and\ \bibinfo {author} {\bibfnamefont {A.~V.}\ \bibnamefont {Zakharov}},\
  }\href {https://doi.org/10.1103/PhysRevD.74.024020} {\bibfield  {journal}
  {\bibinfo  {journal} {Phys. Rev. D}\ }\textbf {\bibinfo {volume} {74}},\
  \bibinfo {pages} {024020} (\bibinfo {year} {2006})},\ \Eprint
  {https://arxiv.org/abs/gr-qc/0602062} {arXiv:gr-qc/0602062} \BibitemShut
  {NoStop}%
\bibitem [{\citenamefont {Ohgami}\ and\ \citenamefont
  {Sakai}(2015)}]{Ohgami:2015nra}%
  \BibitemOpen
  \bibfield  {author} {\bibinfo {author} {\bibfnamefont {T.}~\bibnamefont
  {Ohgami}}\ and\ \bibinfo {author} {\bibfnamefont {N.}~\bibnamefont {Sakai}},\
  }\href {https://doi.org/10.1103/PhysRevD.91.124020} {\bibfield  {journal}
  {\bibinfo  {journal} {Phys. Rev. D}\ }\textbf {\bibinfo {volume} {91}},\
  \bibinfo {pages} {124020} (\bibinfo {year} {2015})},\ \Eprint
  {https://arxiv.org/abs/1704.07065} {arXiv:1704.07065 [gr-qc]} \BibitemShut
  {NoStop}%
\bibitem [{\citenamefont {Cardoso}\ \emph
  {et~al.}(2016{\natexlab{a}})\citenamefont {Cardoso}, \citenamefont
  {Franzin},\ and\ \citenamefont {Pani}}]{Cardoso:2016rao}%
  \BibitemOpen
  \bibfield  {author} {\bibinfo {author} {\bibfnamefont {V.}~\bibnamefont
  {Cardoso}}, \bibinfo {author} {\bibfnamefont {E.}~\bibnamefont {Franzin}},\
  and\ \bibinfo {author} {\bibfnamefont {P.}~\bibnamefont {Pani}},\ }\href
  {https://doi.org/10.1103/PhysRevLett.116.171101} {\bibfield  {journal}
  {\bibinfo  {journal} {Phys. Rev. Lett.}\ }\textbf {\bibinfo {volume} {116}},\
  \bibinfo {pages} {171101} (\bibinfo {year} {2016}{\natexlab{a}})},\ \bibinfo
  {note} {[Erratum: Phys.Rev.Lett. 117, 089902 (2016)]},\ \Eprint
  {https://arxiv.org/abs/1602.07309} {arXiv:1602.07309 [gr-qc]} \BibitemShut
  {NoStop}%
\bibitem [{\citenamefont {Cardoso}\ \emph
  {et~al.}(2016{\natexlab{b}})\citenamefont {Cardoso}, \citenamefont {Hopper},
  \citenamefont {Macedo}, \citenamefont {Palenzuela},\ and\ \citenamefont
  {Pani}}]{Cardoso:2016oxy}%
  \BibitemOpen
  \bibfield  {author} {\bibinfo {author} {\bibfnamefont {V.}~\bibnamefont
  {Cardoso}}, \bibinfo {author} {\bibfnamefont {S.}~\bibnamefont {Hopper}},
  \bibinfo {author} {\bibfnamefont {C.~F.~B.}\ \bibnamefont {Macedo}}, \bibinfo
  {author} {\bibfnamefont {C.}~\bibnamefont {Palenzuela}},\ and\ \bibinfo
  {author} {\bibfnamefont {P.}~\bibnamefont {Pani}},\ }\href
  {https://doi.org/10.1103/PhysRevD.94.084031} {\bibfield  {journal} {\bibinfo
  {journal} {Phys. Rev. D}\ }\textbf {\bibinfo {volume} {94}},\ \bibinfo
  {pages} {084031} (\bibinfo {year} {2016}{\natexlab{b}})},\ \Eprint
  {https://arxiv.org/abs/1608.08637} {arXiv:1608.08637 [gr-qc]} \BibitemShut
  {NoStop}%
\bibitem [{\citenamefont {Wielgus}\ \emph {et~al.}(2020)\citenamefont
  {Wielgus}, \citenamefont {Horak}, \citenamefont {Vincent},\ and\
  \citenamefont {Abramowicz}}]{Wielgus:2020uqz}%
  \BibitemOpen
  \bibfield  {author} {\bibinfo {author} {\bibfnamefont {M.}~\bibnamefont
  {Wielgus}}, \bibinfo {author} {\bibfnamefont {J.}~\bibnamefont {Horak}},
  \bibinfo {author} {\bibfnamefont {F.}~\bibnamefont {Vincent}},\ and\ \bibinfo
  {author} {\bibfnamefont {M.}~\bibnamefont {Abramowicz}},\ }\href
  {https://doi.org/10.1103/PhysRevD.102.084044} {\bibfield  {journal} {\bibinfo
   {journal} {Phys. Rev. D}\ }\textbf {\bibinfo {volume} {102}},\ \bibinfo
  {pages} {084044} (\bibinfo {year} {2020})},\ \Eprint
  {https://arxiv.org/abs/2008.10130} {arXiv:2008.10130 [gr-qc]} \BibitemShut
  {NoStop}%
\bibitem [{\citenamefont {Guerrero}\ \emph {et~al.}(2022)\citenamefont
  {Guerrero}, \citenamefont {Olmo}, \citenamefont {Rubiera-Garcia},\ and\
  \citenamefont {G{\'o}mez}}]{guerrero2022light}%
  \BibitemOpen
  \bibfield  {author} {\bibinfo {author} {\bibfnamefont {M.}~\bibnamefont
  {Guerrero}}, \bibinfo {author} {\bibfnamefont {G.~J.}\ \bibnamefont {Olmo}},
  \bibinfo {author} {\bibfnamefont {D.}~\bibnamefont {Rubiera-Garcia}},\ and\
  \bibinfo {author} {\bibfnamefont {D.~S.-C.}\ \bibnamefont {G{\'o}mez}},\
  }\href@noop {} {\bibfield  {journal} {\bibinfo  {journal} {Physical Review
  D}\ }\textbf {\bibinfo {volume} {105}},\ \bibinfo {pages} {084057} (\bibinfo
  {year} {2022})}\BibitemShut {NoStop}%
\bibitem [{\citenamefont {Calder\'on~Bustillo}\ \emph
  {et~al.}(2021)\citenamefont {Calder\'on~Bustillo}, \citenamefont
  {Sanchis-Gual}, \citenamefont {Torres-Forn\'e}, \citenamefont {Font},
  \citenamefont {Vajpeyi}, \citenamefont {Smith}, \citenamefont {Herdeiro},
  \citenamefont {Radu},\ and\ \citenamefont
  {Leong}}]{CalderonBustillo:2020fyi}%
  \BibitemOpen
  \bibfield  {author} {\bibinfo {author} {\bibfnamefont {J.}~\bibnamefont
  {Calder\'on~Bustillo}}, \bibinfo {author} {\bibfnamefont {N.}~\bibnamefont
  {Sanchis-Gual}}, \bibinfo {author} {\bibfnamefont {A.}~\bibnamefont
  {Torres-Forn\'e}}, \bibinfo {author} {\bibfnamefont {J.~A.}\ \bibnamefont
  {Font}}, \bibinfo {author} {\bibfnamefont {A.}~\bibnamefont {Vajpeyi}},
  \bibinfo {author} {\bibfnamefont {R.}~\bibnamefont {Smith}}, \bibinfo
  {author} {\bibfnamefont {C.}~\bibnamefont {Herdeiro}}, \bibinfo {author}
  {\bibfnamefont {E.}~\bibnamefont {Radu}},\ and\ \bibinfo {author}
  {\bibfnamefont {S.~H.~W.}\ \bibnamefont {Leong}},\ }\href
  {https://doi.org/10.1103/PhysRevLett.126.081101} {\bibfield  {journal}
  {\bibinfo  {journal} {Phys. Rev. Lett.}\ }\textbf {\bibinfo {volume} {126}},\
  \bibinfo {pages} {081101} (\bibinfo {year} {2021})},\ \Eprint
  {https://arxiv.org/abs/2009.05376} {arXiv:2009.05376 [gr-qc]} \BibitemShut
  {NoStop}%
\bibitem [{\citenamefont {Cardoso}\ and\ \citenamefont
  {Pani}(2019)}]{Cardoso:2019rvt}%
  \BibitemOpen
  \bibfield  {author} {\bibinfo {author} {\bibfnamefont {V.}~\bibnamefont
  {Cardoso}}\ and\ \bibinfo {author} {\bibfnamefont {P.}~\bibnamefont {Pani}},\
  }\href {https://doi.org/10.1007/s41114-019-0020-4} {\bibfield  {journal}
  {\bibinfo  {journal} {Living Rev. Rel.}\ }\textbf {\bibinfo {volume} {22}},\
  \bibinfo {pages} {4} (\bibinfo {year} {2019})},\ \Eprint
  {https://arxiv.org/abs/1904.05363} {arXiv:1904.05363 [gr-qc]} \BibitemShut
  {NoStop}%
\bibitem [{\citenamefont {Visser}\ and\ \citenamefont
  {Hochberg}(1997)}]{Visser:1997yn}%
  \BibitemOpen
  \bibfield  {author} {\bibinfo {author} {\bibfnamefont {M.}~\bibnamefont
  {Visser}}\ and\ \bibinfo {author} {\bibfnamefont {D.}~\bibnamefont
  {Hochberg}},\ }\href@noop {} {\bibfield  {journal} {\bibinfo  {journal}
  {Annals Israel Phys. Soc.}\ }\textbf {\bibinfo {volume} {13}},\ \bibinfo
  {pages} {249} (\bibinfo {year} {1997})},\ \Eprint
  {https://arxiv.org/abs/gr-qc/9710001} {arXiv:gr-qc/9710001} \BibitemShut
  {NoStop}%
\bibitem [{\citenamefont {Mas\'o-Ferrando}\ \emph
  {et~al.}(2023{\natexlab{a}})\citenamefont {Mas\'o-Ferrando}, \citenamefont
  {Sanchis-Gual}, \citenamefont {Font},\ and\ \citenamefont
  {Olmo}}]{Maso-Ferrando:2023nju}%
  \BibitemOpen
  \bibfield  {author} {\bibinfo {author} {\bibfnamefont {A.}~\bibnamefont
  {Mas\'o-Ferrando}}, \bibinfo {author} {\bibfnamefont {N.}~\bibnamefont
  {Sanchis-Gual}}, \bibinfo {author} {\bibfnamefont {J.~A.}\ \bibnamefont
  {Font}},\ and\ \bibinfo {author} {\bibfnamefont {G.~J.}\ \bibnamefont
  {Olmo}},\ }\href {https://doi.org/10.1088/1475-7516/2023/06/028} {\bibfield
  {journal} {\bibinfo  {journal} {JCAP}\ }\textbf {\bibinfo {volume} {06}},\
  \bibinfo {pages} {028}},\ \Eprint {https://arxiv.org/abs/2304.12018}
  {arXiv:2304.12018 [gr-qc]} \BibitemShut {NoStop}%
\bibitem [{\citenamefont {Mas\'o-Ferrando}\ \emph
  {et~al.}(2023{\natexlab{b}})\citenamefont {Mas\'o-Ferrando}, \citenamefont
  {Sanchis-Gual}, \citenamefont {Font},\ and\ \citenamefont
  {Olmo}}]{Maso-Ferrando:2023wtz}%
  \BibitemOpen
  \bibfield  {author} {\bibinfo {author} {\bibfnamefont {A.}~\bibnamefont
  {Mas\'o-Ferrando}}, \bibinfo {author} {\bibfnamefont {N.}~\bibnamefont
  {Sanchis-Gual}}, \bibinfo {author} {\bibfnamefont {J.~A.}\ \bibnamefont
  {Font}},\ and\ \bibinfo {author} {\bibfnamefont {G.~J.}\ \bibnamefont
  {Olmo}},\ }\href@noop {} {\  (\bibinfo {year} {2023}{\natexlab{b}})},\
  \Eprint {https://arxiv.org/abs/2309.14912} {arXiv:2309.14912 [gr-qc]}
  \BibitemShut {NoStop}%
\bibitem [{\citenamefont {Ellis}(1973)}]{ellis1973ether}%
  \BibitemOpen
  \bibfield  {author} {\bibinfo {author} {\bibfnamefont {H.~G.}\ \bibnamefont
  {Ellis}},\ }\href@noop {} {\bibfield  {journal} {\bibinfo  {journal} {Journal
  of Mathematical Physics}\ }\textbf {\bibinfo {volume} {14}},\ \bibinfo
  {pages} {104} (\bibinfo {year} {1973})}\BibitemShut {NoStop}%
\bibitem [{\citenamefont {Magalh\~aes}\ \emph {et~al.}()\citenamefont
  {Magalh\~aes}, \citenamefont {Mas{\'o}-Ferrando}, \citenamefont {Bombacigno},
  \citenamefont {Olmo},\ and\ \citenamefont {B.~Crispino}}]{iamthefuture}%
  \BibitemOpen
  \bibfield  {author} {\bibinfo {author} {\bibfnamefont {R.~B.}\ \bibnamefont
  {Magalh\~aes}}, \bibinfo {author} {\bibfnamefont {A.}~\bibnamefont
  {Mas{\'o}-Ferrando}}, \bibinfo {author} {\bibfnamefont {F.}~\bibnamefont
  {Bombacigno}}, \bibinfo {author} {\bibfnamefont {G.~J.}\ \bibnamefont
  {Olmo}},\ and\ \bibinfo {author} {\bibfnamefont {L.~C.}\ \bibnamefont
  {B.~Crispino}},\ }\href@noop {} {\ }\Eprint {https://arxiv.org/abs/in
  preparation} {in preparation} \BibitemShut {NoStop}%
\bibitem [{\citenamefont {Cardoso}\ \emph
  {et~al.}(2016{\natexlab{c}})\citenamefont {Cardoso}, \citenamefont {Hopper},
  \citenamefont {Macedo}, \citenamefont {Palenzuela},\ and\ \citenamefont
  {Pani}}]{cardoso2016gravitational}%
  \BibitemOpen
  \bibfield  {author} {\bibinfo {author} {\bibfnamefont {V.}~\bibnamefont
  {Cardoso}}, \bibinfo {author} {\bibfnamefont {S.}~\bibnamefont {Hopper}},
  \bibinfo {author} {\bibfnamefont {C.~F.}\ \bibnamefont {Macedo}}, \bibinfo
  {author} {\bibfnamefont {C.}~\bibnamefont {Palenzuela}},\ and\ \bibinfo
  {author} {\bibfnamefont {P.}~\bibnamefont {Pani}},\ }\href@noop {} {\bibfield
   {journal} {\bibinfo  {journal} {Physical review D}\ }\textbf {\bibinfo
  {volume} {94}},\ \bibinfo {pages} {084031} (\bibinfo {year}
  {2016}{\natexlab{c}})}\BibitemShut {NoStop}%
\bibitem [{\citenamefont {Cardoso}\ and\ \citenamefont
  {Pani}(2017)}]{cardoso2017tests}%
  \BibitemOpen
  \bibfield  {author} {\bibinfo {author} {\bibfnamefont {V.}~\bibnamefont
  {Cardoso}}\ and\ \bibinfo {author} {\bibfnamefont {P.}~\bibnamefont {Pani}},\
  }\href@noop {} {\bibfield  {journal} {\bibinfo  {journal} {Nature Astronomy}\
  }\textbf {\bibinfo {volume} {1}},\ \bibinfo {pages} {586} (\bibinfo {year}
  {2017})}\BibitemShut {NoStop}%
\bibitem [{\citenamefont {Bueno}\ \emph {et~al.}(2018)\citenamefont {Bueno},
  \citenamefont {Cano}, \citenamefont {Goelen}, \citenamefont {Hertog},\ and\
  \citenamefont {Vercnocke}}]{bueno2018echoes}%
  \BibitemOpen
  \bibfield  {author} {\bibinfo {author} {\bibfnamefont {P.}~\bibnamefont
  {Bueno}}, \bibinfo {author} {\bibfnamefont {P.~A.}\ \bibnamefont {Cano}},
  \bibinfo {author} {\bibfnamefont {F.}~\bibnamefont {Goelen}}, \bibinfo
  {author} {\bibfnamefont {T.}~\bibnamefont {Hertog}},\ and\ \bibinfo {author}
  {\bibfnamefont {B.}~\bibnamefont {Vercnocke}},\ }\href@noop {} {\bibfield
  {journal} {\bibinfo  {journal} {Physical Review D}\ }\textbf {\bibinfo
  {volume} {97}},\ \bibinfo {pages} {024040} (\bibinfo {year}
  {2018})}\BibitemShut {NoStop}%
\bibitem [{\citenamefont {Bronnikov}\ and\ \citenamefont
  {Konoplya}(2020)}]{bronnikov2020echoes}%
  \BibitemOpen
  \bibfield  {author} {\bibinfo {author} {\bibfnamefont {K.~A.}\ \bibnamefont
  {Bronnikov}}\ and\ \bibinfo {author} {\bibfnamefont {R.~A.}\ \bibnamefont
  {Konoplya}},\ }\href@noop {} {\bibfield  {journal} {\bibinfo  {journal}
  {Physical Review D}\ }\textbf {\bibinfo {volume} {101}},\ \bibinfo {pages}
  {064004} (\bibinfo {year} {2020})}\BibitemShut {NoStop}%
\bibitem [{\citenamefont {Churilova}\ \emph {et~al.}(2021)\citenamefont
  {Churilova}, \citenamefont {Konoplya}, \citenamefont {Stuchlik},\ and\
  \citenamefont {Zhidenko}}]{churilova2021wormholes}%
  \BibitemOpen
  \bibfield  {author} {\bibinfo {author} {\bibfnamefont {M.}~\bibnamefont
  {Churilova}}, \bibinfo {author} {\bibfnamefont {R.}~\bibnamefont {Konoplya}},
  \bibinfo {author} {\bibfnamefont {Z.}~\bibnamefont {Stuchlik}},\ and\
  \bibinfo {author} {\bibfnamefont {A.}~\bibnamefont {Zhidenko}},\ }\href@noop
  {} {\bibfield  {journal} {\bibinfo  {journal} {Journal of Cosmology and
  Astroparticle Physics}\ }\textbf {\bibinfo {volume} {2021}}\bibinfo  {number}
  { (10)},\ \bibinfo {pages} {010}}\BibitemShut {NoStop}%
\bibitem [{\citenamefont {Yang}\ \emph {et~al.}(2021)\citenamefont {Yang},
  \citenamefont {Liu}, \citenamefont {Xu}, \citenamefont {Xing}, \citenamefont
  {Wu},\ and\ \citenamefont {Long}}]{yang2021echoes}%
  \BibitemOpen
\bibfield  {number} {  }\bibfield  {author} {\bibinfo {author} {\bibfnamefont
  {Y.}~\bibnamefont {Yang}}, \bibinfo {author} {\bibfnamefont {D.}~\bibnamefont
  {Liu}}, \bibinfo {author} {\bibfnamefont {Z.}~\bibnamefont {Xu}}, \bibinfo
  {author} {\bibfnamefont {Y.}~\bibnamefont {Xing}}, \bibinfo {author}
  {\bibfnamefont {S.}~\bibnamefont {Wu}},\ and\ \bibinfo {author}
  {\bibfnamefont {Z.-W.}\ \bibnamefont {Long}},\ }\href@noop {} {\bibfield
  {journal} {\bibinfo  {journal} {Physical Review D}\ }\textbf {\bibinfo
  {volume} {104}},\ \bibinfo {pages} {104021} (\bibinfo {year}
  {2021})}\BibitemShut {NoStop}%
\bibitem [{\citenamefont {Afonso}\ \emph {et~al.}(2019)\citenamefont {Afonso},
  \citenamefont {Olmo}, \citenamefont {Orazi},\ and\ \citenamefont
  {Rubiera-Garcia}}]{afonso2019new}%
  \BibitemOpen
  \bibfield  {author} {\bibinfo {author} {\bibfnamefont {V.~I.}\ \bibnamefont
  {Afonso}}, \bibinfo {author} {\bibfnamefont {G.~J.}\ \bibnamefont {Olmo}},
  \bibinfo {author} {\bibfnamefont {E.}~\bibnamefont {Orazi}},\ and\ \bibinfo
  {author} {\bibfnamefont {D.}~\bibnamefont {Rubiera-Garcia}},\ }\href@noop {}
  {\bibfield  {journal} {\bibinfo  {journal} {Journal of Cosmology and
  Astroparticle Physics}\ }\textbf {\bibinfo {volume} {2019}}\bibinfo  {number}
  { (12)},\ \bibinfo {pages} {044}}\BibitemShut {NoStop}%
\bibitem [{\citenamefont {Magalh{\~a}es}\ \emph {et~al.}(2022)\citenamefont
  {Magalh{\~a}es}, \citenamefont {Crispino},\ and\ \citenamefont
  {Olmo}}]{magalhaes2022compact}%
  \BibitemOpen
\bibfield  {number} {  }\bibfield  {author} {\bibinfo {author} {\bibfnamefont
  {R.~B.}\ \bibnamefont {Magalh{\~a}es}}, \bibinfo {author} {\bibfnamefont
  {L.~C.}\ \bibnamefont {Crispino}},\ and\ \bibinfo {author} {\bibfnamefont
  {G.~J.}\ \bibnamefont {Olmo}},\ }\href@noop {} {\bibfield  {journal}
  {\bibinfo  {journal} {Physical Review D}\ }\textbf {\bibinfo {volume}
  {105}},\ \bibinfo {pages} {064007} (\bibinfo {year} {2022})}\BibitemShut
  {NoStop}%
\bibitem [{\citenamefont {Sharma}\ and\ \citenamefont
  {Ghosh}(2021)}]{sharma2021generalised}%
  \BibitemOpen
  \bibfield  {author} {\bibinfo {author} {\bibfnamefont {V.}~\bibnamefont
  {Sharma}}\ and\ \bibinfo {author} {\bibfnamefont {S.}~\bibnamefont {Ghosh}},\
  }\href@noop {} {\bibfield  {journal} {\bibinfo  {journal} {The European
  Physical Journal C}\ }\textbf {\bibinfo {volume} {81}},\ \bibinfo {pages}
  {1004} (\bibinfo {year} {2021})}\BibitemShut {NoStop}%
\bibitem [{\citenamefont {Capozziello}\ \emph {et~al.}(2012)\citenamefont
  {Capozziello}, \citenamefont {Harko}, \citenamefont {Koivisto}, \citenamefont
  {Lobo},\ and\ \citenamefont {Olmo}}]{Capozziello:2012hr}%
  \BibitemOpen
  \bibfield  {author} {\bibinfo {author} {\bibfnamefont {S.}~\bibnamefont
  {Capozziello}}, \bibinfo {author} {\bibfnamefont {T.}~\bibnamefont {Harko}},
  \bibinfo {author} {\bibfnamefont {T.~S.}\ \bibnamefont {Koivisto}}, \bibinfo
  {author} {\bibfnamefont {F.~S.~N.}\ \bibnamefont {Lobo}},\ and\ \bibinfo
  {author} {\bibfnamefont {G.~J.}\ \bibnamefont {Olmo}},\ }\href
  {https://doi.org/10.1103/PhysRevD.86.127504} {\bibfield  {journal} {\bibinfo
  {journal} {Phys. Rev. D}\ }\textbf {\bibinfo {volume} {86}},\ \bibinfo
  {pages} {127504} (\bibinfo {year} {2012})},\ \Eprint
  {https://arxiv.org/abs/1209.5862} {arXiv:1209.5862 [gr-qc]} \BibitemShut
  {NoStop}%
\bibitem [{\citenamefont {Guendelman}\ \emph {et~al.}(2013)\citenamefont
  {Guendelman}, \citenamefont {Olmo}, \citenamefont {Rubiera-Garcia},\ and\
  \citenamefont {Vasihoun}}]{Guendelman:2013sca}%
  \BibitemOpen
  \bibfield  {author} {\bibinfo {author} {\bibfnamefont {E.~I.}\ \bibnamefont
  {Guendelman}}, \bibinfo {author} {\bibfnamefont {G.~J.}\ \bibnamefont
  {Olmo}}, \bibinfo {author} {\bibfnamefont {D.}~\bibnamefont
  {Rubiera-Garcia}},\ and\ \bibinfo {author} {\bibfnamefont {M.}~\bibnamefont
  {Vasihoun}},\ }\href {https://doi.org/10.1016/j.physletb.2013.09.039}
  {\bibfield  {journal} {\bibinfo  {journal} {Phys. Lett. B}\ }\textbf
  {\bibinfo {volume} {726}},\ \bibinfo {pages} {870} (\bibinfo {year}
  {2013})},\ \Eprint {https://arxiv.org/abs/1306.6769} {arXiv:1306.6769
  [hep-th]} \BibitemShut {NoStop}%
\bibitem [{\citenamefont {Bambi}\ \emph {et~al.}(2016)\citenamefont {Bambi},
  \citenamefont {Cardenas-Avendano}, \citenamefont {Olmo},\ and\ \citenamefont
  {Rubiera-Garcia}}]{Bambi:2015zch}%
  \BibitemOpen
  \bibfield  {author} {\bibinfo {author} {\bibfnamefont {C.}~\bibnamefont
  {Bambi}}, \bibinfo {author} {\bibfnamefont {A.}~\bibnamefont
  {Cardenas-Avendano}}, \bibinfo {author} {\bibfnamefont {G.~J.}\ \bibnamefont
  {Olmo}},\ and\ \bibinfo {author} {\bibfnamefont {D.}~\bibnamefont
  {Rubiera-Garcia}},\ }\href {https://doi.org/10.1103/PhysRevD.93.064016}
  {\bibfield  {journal} {\bibinfo  {journal} {Phys. Rev. D}\ }\textbf {\bibinfo
  {volume} {93}},\ \bibinfo {pages} {064016} (\bibinfo {year} {2016})},\
  \Eprint {https://arxiv.org/abs/1511.03755} {arXiv:1511.03755 [gr-qc]}
  \BibitemShut {NoStop}%
\bibitem [{\citenamefont {\"Ovg\"un}\ \emph {et~al.}(2019)\citenamefont
  {\"Ovg\"un}, \citenamefont {Jusufi},\ and\ \citenamefont
  {Sakall\i{}}}]{Ovgun:2018xys}%
  \BibitemOpen
  \bibfield  {author} {\bibinfo {author} {\bibfnamefont {A.}~\bibnamefont
  {\"Ovg\"un}}, \bibinfo {author} {\bibfnamefont {K.}~\bibnamefont {Jusufi}},\
  and\ \bibinfo {author} {\bibfnamefont {I.}~\bibnamefont {Sakall\i{}}},\
  }\href {https://doi.org/10.1103/PhysRevD.99.024042} {\bibfield  {journal}
  {\bibinfo  {journal} {Phys. Rev. D}\ }\textbf {\bibinfo {volume} {99}},\
  \bibinfo {pages} {024042} (\bibinfo {year} {2019})},\ \Eprint
  {https://arxiv.org/abs/1804.09911} {arXiv:1804.09911 [gr-qc]} \BibitemShut
  {NoStop}%
\bibitem [{\citenamefont {Nascimento}\ \emph {et~al.}(2019)\citenamefont
  {Nascimento}, \citenamefont {Olmo}, \citenamefont {Porfirio}, \citenamefont
  {Petrov},\ and\ \citenamefont {Soares}}]{Nascimento:2018sir}%
  \BibitemOpen
  \bibfield  {author} {\bibinfo {author} {\bibfnamefont {J.~R.}\ \bibnamefont
  {Nascimento}}, \bibinfo {author} {\bibfnamefont {G.~J.}\ \bibnamefont
  {Olmo}}, \bibinfo {author} {\bibfnamefont {P.~J.}\ \bibnamefont {Porfirio}},
  \bibinfo {author} {\bibfnamefont {A.~Y.}\ \bibnamefont {Petrov}},\ and\
  \bibinfo {author} {\bibfnamefont {A.~R.}\ \bibnamefont {Soares}},\ }\href
  {https://doi.org/10.1103/PhysRevD.99.064053} {\bibfield  {journal} {\bibinfo
  {journal} {Phys. Rev. D}\ }\textbf {\bibinfo {volume} {99}},\ \bibinfo
  {pages} {064053} (\bibinfo {year} {2019})},\ \Eprint
  {https://arxiv.org/abs/1812.00471} {arXiv:1812.00471 [gr-qc]} \BibitemShut
  {NoStop}%
\bibitem [{\citenamefont {Rosa}\ \emph {et~al.}(2018)\citenamefont {Rosa},
  \citenamefont {Lemos},\ and\ \citenamefont {Lobo}}]{Rosa:2018jwp}%
  \BibitemOpen
  \bibfield  {author} {\bibinfo {author} {\bibfnamefont {J.~a.~L.}\
  \bibnamefont {Rosa}}, \bibinfo {author} {\bibfnamefont {J.~P.~S.}\
  \bibnamefont {Lemos}},\ and\ \bibinfo {author} {\bibfnamefont {F.~S.~N.}\
  \bibnamefont {Lobo}},\ }\href {https://doi.org/10.1103/PhysRevD.98.064054}
  {\bibfield  {journal} {\bibinfo  {journal} {Phys. Rev. D}\ }\textbf {\bibinfo
  {volume} {98}},\ \bibinfo {pages} {064054} (\bibinfo {year} {2018})},\
  \Eprint {https://arxiv.org/abs/1808.08975} {arXiv:1808.08975 [gr-qc]}
  \BibitemShut {NoStop}%
\bibitem [{\citenamefont {Fu}\ \emph {et~al.}(2019)\citenamefont {Fu},
  \citenamefont {Grado-White},\ and\ \citenamefont {Marolf}}]{Fu:2018oaq}%
  \BibitemOpen
  \bibfield  {author} {\bibinfo {author} {\bibfnamefont {Z.}~\bibnamefont
  {Fu}}, \bibinfo {author} {\bibfnamefont {B.}~\bibnamefont {Grado-White}},\
  and\ \bibinfo {author} {\bibfnamefont {D.}~\bibnamefont {Marolf}},\ }\href
  {https://doi.org/10.1088/1361-6382/aafcea} {\bibfield  {journal} {\bibinfo
  {journal} {Class. Quant. Grav.}\ }\textbf {\bibinfo {volume} {36}},\ \bibinfo
  {pages} {045006} (\bibinfo {year} {2019})},\ \bibinfo {note} {[Erratum:
  Class.Quant.Grav. 36, 249501 (2019)]},\ \Eprint
  {https://arxiv.org/abs/1807.07917} {arXiv:1807.07917 [hep-th]} \BibitemShut
  {NoStop}%
\bibitem [{\citenamefont {Bronnikov}(2019)}]{Bronnikov:2019ugl}%
  \BibitemOpen
  \bibfield  {author} {\bibinfo {author} {\bibfnamefont {K.~A.}\ \bibnamefont
  {Bronnikov}},\ }\href {https://doi.org/10.1134/S0202289319040030} {\bibfield
  {journal} {\bibinfo  {journal} {Grav. Cosmol.}\ }\textbf {\bibinfo {volume}
  {25}},\ \bibinfo {pages} {331} (\bibinfo {year} {2019})},\ \Eprint
  {https://arxiv.org/abs/1908.02012} {arXiv:1908.02012 [gr-qc]} \BibitemShut
  {NoStop}%
\bibitem [{\citenamefont {Magalh\~aes}\ \emph {et~al.}(2023)\citenamefont
  {Magalh\~aes}, \citenamefont {Mas\'o-Ferrando}, \citenamefont {Olmo},\ and\
  \citenamefont {Crispino}}]{PhysRevD.108.024063}%
  \BibitemOpen
  \bibfield  {author} {\bibinfo {author} {\bibfnamefont {R.~B.}\ \bibnamefont
  {Magalh\~aes}}, \bibinfo {author} {\bibfnamefont {A.}~\bibnamefont
  {Mas\'o-Ferrando}}, \bibinfo {author} {\bibfnamefont {G.~J.}\ \bibnamefont
  {Olmo}},\ and\ \bibinfo {author} {\bibfnamefont {L.~C.~B.}\ \bibnamefont
  {Crispino}},\ }\href {https://doi.org/10.1103/PhysRevD.108.024063} {\bibfield
   {journal} {\bibinfo  {journal} {Phys. Rev. D}\ }\textbf {\bibinfo {volume}
  {108}},\ \bibinfo {pages} {024063} (\bibinfo {year} {2023})}\BibitemShut
  {NoStop}%
\bibitem [{\citenamefont {Visser}\ and\ \citenamefont
  {Barcelo}(2000)}]{visser2000energy}%
  \BibitemOpen
  \bibfield  {author} {\bibinfo {author} {\bibfnamefont {M.}~\bibnamefont
  {Visser}}\ and\ \bibinfo {author} {\bibfnamefont {C.}~\bibnamefont
  {Barcelo}},\ }in\ \href@noop {} {\emph {\bibinfo {booktitle} {Cosmo-99}}}\
  (\bibinfo  {publisher} {World Scientific},\ \bibinfo {year} {2000})\ pp.\
  \bibinfo {pages} {98--112}\BibitemShut {NoStop}%
\bibitem [{\citenamefont {Curiel}(2017)}]{curiel2017primer}%
  \BibitemOpen
  \bibfield  {author} {\bibinfo {author} {\bibfnamefont {E.}~\bibnamefont
  {Curiel}},\ }\href@noop {} {\bibfield  {journal} {\bibinfo  {journal}
  {Towards a theory of spacetime theories}\ ,\ \bibinfo {pages} {43}} (\bibinfo
  {year} {2017})}\BibitemShut {NoStop}%
\bibitem [{\citenamefont {Gundlach}\ \emph {et~al.}(1994)\citenamefont
  {Gundlach}, \citenamefont {Price},\ and\ \citenamefont
  {Pullin}}]{Gundlach:1993tp}%
  \BibitemOpen
  \bibfield  {author} {\bibinfo {author} {\bibfnamefont {C.}~\bibnamefont
  {Gundlach}}, \bibinfo {author} {\bibfnamefont {R.~H.}\ \bibnamefont
  {Price}},\ and\ \bibinfo {author} {\bibfnamefont {J.}~\bibnamefont
  {Pullin}},\ }\href {https://doi.org/10.1103/PhysRevD.49.883} {\bibfield
  {journal} {\bibinfo  {journal} {Phys. Rev. D}\ }\textbf {\bibinfo {volume}
  {49}},\ \bibinfo {pages} {883} (\bibinfo {year} {1994})},\ \Eprint
  {https://arxiv.org/abs/gr-qc/9307009} {arXiv:gr-qc/9307009} \BibitemShut
  {NoStop}%
\bibitem [{\citenamefont {Kokkotas}\ \emph {et~al.}(2011)\citenamefont
  {Kokkotas}, \citenamefont {Konoplya},\ and\ \citenamefont
  {Zhidenko}}]{kokkotas2011quasinormal}%
  \BibitemOpen
  \bibfield  {author} {\bibinfo {author} {\bibfnamefont {K.}~\bibnamefont
  {Kokkotas}}, \bibinfo {author} {\bibfnamefont {R.}~\bibnamefont {Konoplya}},\
  and\ \bibinfo {author} {\bibfnamefont {A.}~\bibnamefont {Zhidenko}},\
  }\href@noop {} {\bibfield  {journal} {\bibinfo  {journal} {Physical Review
  D}\ }\textbf {\bibinfo {volume} {83}},\ \bibinfo {pages} {024031} (\bibinfo
  {year} {2011})}\BibitemShut {NoStop}%
\bibitem [{\citenamefont {Turimov}\ \emph {et~al.}(2019)\citenamefont
  {Turimov}, \citenamefont {Toshmatov}, \citenamefont {Ahmedov},\ and\
  \citenamefont {Stuchl{\'\i}k}}]{turimov2019quasinormal}%
  \BibitemOpen
  \bibfield  {author} {\bibinfo {author} {\bibfnamefont {B.}~\bibnamefont
  {Turimov}}, \bibinfo {author} {\bibfnamefont {B.}~\bibnamefont {Toshmatov}},
  \bibinfo {author} {\bibfnamefont {B.}~\bibnamefont {Ahmedov}},\ and\ \bibinfo
  {author} {\bibfnamefont {Z.}~\bibnamefont {Stuchl{\'\i}k}},\ }\href@noop {}
  {\bibfield  {journal} {\bibinfo  {journal} {Physical Review D}\ }\textbf
  {\bibinfo {volume} {100}},\ \bibinfo {pages} {084038} (\bibinfo {year}
  {2019})}\BibitemShut {NoStop}%
\bibitem [{\citenamefont {Ohashi}\ and\ \citenamefont
  {Sakagami}(2004)}]{ohashi2004massive}%
  \BibitemOpen
  \bibfield  {author} {\bibinfo {author} {\bibfnamefont {A.}~\bibnamefont
  {Ohashi}}\ and\ \bibinfo {author} {\bibfnamefont {M.-a.}\ \bibnamefont
  {Sakagami}},\ }\href@noop {} {\bibfield  {journal} {\bibinfo  {journal}
  {Classical and Quantum Gravity}\ }\textbf {\bibinfo {volume} {21}},\ \bibinfo
  {pages} {3973} (\bibinfo {year} {2004})}\BibitemShut {NoStop}%
\bibitem [{\citenamefont {Konoplya}\ and\ \citenamefont
  {Zhidenko}(2005)}]{konoplya2005decay}%
  \BibitemOpen
  \bibfield  {author} {\bibinfo {author} {\bibfnamefont {R.~A.}\ \bibnamefont
  {Konoplya}}\ and\ \bibinfo {author} {\bibfnamefont {A.}~\bibnamefont
  {Zhidenko}},\ }\href@noop {} {\bibfield  {journal} {\bibinfo  {journal}
  {Physics Letters B}\ }\textbf {\bibinfo {volume} {609}},\ \bibinfo {pages}
  {377} (\bibinfo {year} {2005})}\BibitemShut {NoStop}%
\bibitem [{\citenamefont {Shao}\ \emph {et~al.}(2022)\citenamefont {Shao},
  \citenamefont {Tan}, \citenamefont {Shao}, \citenamefont {Lin},\ and\
  \citenamefont {Qian}}]{shao2022quasinormal}%
  \BibitemOpen
  \bibfield  {author} {\bibinfo {author} {\bibfnamefont {C.-Y.}\ \bibnamefont
  {Shao}}, \bibinfo {author} {\bibfnamefont {Y.-J.}\ \bibnamefont {Tan}},
  \bibinfo {author} {\bibfnamefont {C.-G.}\ \bibnamefont {Shao}}, \bibinfo
  {author} {\bibfnamefont {K.}~\bibnamefont {Lin}},\ and\ \bibinfo {author}
  {\bibfnamefont {W.-L.}\ \bibnamefont {Qian}},\ }\href@noop {} {\bibfield
  {journal} {\bibinfo  {journal} {Chinese Physics C}\ }\textbf {\bibinfo
  {volume} {46}},\ \bibinfo {pages} {105103} (\bibinfo {year}
  {2022})}\BibitemShut {NoStop}%
\end{thebibliography}%

\end{document}